\documentstyle[epsf,eqsecnum,prc,aps]{revtex}

\begin{document}
\draft
\twocolumn[\columnwidth\textwidth\csname@twocolumnfalse\endcsname
\title{Full $pf$ shell study of $A=47$ and $A=49$ nuclei}
\author{G. Mart\'{\i}nez-Pinedo$^{1}$, A. P. Zuker$^{1,2}$,
A. Poves$^{1}$ and E. Caurier$^{2}$}

\address{$^{1}$ Departamento de F\'{\i}sica Te\'orica C-XI, Universidad
  Aut\'onoma de Madrid, E--28049 Madrid, Spain}

\address{$^{2}$Groupe de Physique Th\'eorique, Centre de Recherches
  Nucl\'eaires ( IN2P3-CNRS-Universit\'e Louis Pasteur) B\^at.40/1,
 F--67037 Strasbourg Cedex~2, France}
\date{\today}
\maketitle
\begin{abstract}
  Complete diagonalizations in the $pf$ major shell, lead to very good
  agreement with the experimental data (level schemes, transitions
  rates, and static moments) for the $A=47$ and $A=49$ isotopes of Ca,
  Sc, Ti, V, Cr, and Mn. Gamow-Teller and M1 strength functions are
  calculated. The necessary monopole modifications to the realistic
  interactions are shown to be critically tested by the spectroscopic
  factors for one particle transfer from $^{48}$Ca, reproduced in
  detail by the calculations.  The collective behaviour of $^{47}$Ti,
  and of the mirror pairs $^{47}$V--$^{47}$Cr and $^{49}$Cr--$^{49}$Mn
  is found to follow at low spins the particle plus rotor model. It is
  then analysed in terms of the approximate quasi-SU(3) symmetry, for
  which some new results are given.
\end{abstract}
\addvspace{5mm}
]

\section{INTRODUCTION}

In a recent paper~\cite{a48} hereafter referred as I, the A=48 nuclei
were studied through exact diagonalizations in the full $pf$ shell,
using a realistic interaction whose monopole centroids had been
minimally modified to cure the bad saturation properties
characteristic of all forces that describe adequately the NN phase
shifts. Close agreement with experiment was obtained for the
observables (level schemes, static moments and electromagnetic
transitions) near the ground states, and at higher energies as shown
by Gamow Teller strength functions calculated for A=48 and other
nuclei in the region~\cite{quench,fierros}.

One of the most interesting findings in~\cite{a48} was the backbending
behaviour of $^{48}$Cr, simultaneously established
experimentally~\cite{cameron}, and reminiscent of energy patterns
hitherto found only in much heavier rotational nuclei.  In
ref.~\cite{cr48} the question was treated in some detail, establishing
the equivalence of shell model and mean field treatments, and a
similar study is now avilable for $^{50}$Cr~\cite{cr50}.  Recent
experiments indicate that the agreement with the calculated yrast
energies in $^{48}$Cr has become almost perfect, and further data on
the collective properties in the A=46-50 region are
forthcoming~\cite{silvia}.

\smallskip

It appears that an exhaustive study of the A=47 and A=49 nuclei
along the lines of I, and using the same interaction, is a natural
next step. This paper is devoted to it, and it is organized so as to
address three basic issues:
\begin{description}
\item [Detailed spectroscopy] A much decried aspect of Nuclear
  Physics, often referred to as ``zoology'', but nevertheless an
  essential test of the quality of the calculations. This test is
notoriously more difficult to pass in odd nuclei.
\item [The interaction] It is of interest to illustrate why it is
  sufficient to study closed shell nuclei, and the single-particle and
  single-hole excitations based on them, to obtain
   the monopole modifications that transform the realistic
  interactions into successful ones.
\item [Collective properties] The behaviour of yrast lines in
  rotor-like nuclei did not seem amenable to exact shell model
  diagonalizations. Since the A=46-50 region offers a counter example,
  it is well worth going into it in some detail.
\end{description}

Accordingly, we shall proceed as follows:
 
Section~\ref{sec:spectro} will be devoted to detailed spectroscopy by
comparing calculations with the impressive body of data in Burrows's
compilation~\cite{burrows_47,burrows_49,NDS}. The agreement turns out
to be particularly satisfactory when the experimental results are
unambiguous, suggesting that the calculations become useful guides in
the cases of difficult assignments.

Section~\ref{sec:m1gt} deals with Gamow Teller and M1 strength
functions. Though there are no measures of these quantities some of
them may be of special interest. In particular $^{47}$Ti is predicted
to exhibit a strong ``scissors''-like excitation.

In section~\ref{sec:specfac} we analyze the spectroscopic factors for
one particle transfer from $^{48}$Ca to $^{49}$Ca and $^{49}$Sc, and
explain their direct bearing on the minimal monopole modifications
that cure the problems of the Kuo Brown interaction (KB)~\cite{KB},
transforming it into the successful variant that we use
(KB3)~\cite{pasquini,pasquiniz,pozu1,pozu2}.

Rotational properties of the two mirror pairs obtained by adding or
removing a particle from $^{48}$Cr will be studied in
section~\ref{sec:collective}. It will be seen that several low lying
bands are almost perfectly described---up to some critical spin---by
strong particle (or hole)-rotor couplings. The microscopic mechanisms
at the origin of the collective behaviour can be understood in terms
of quasi-SU(3)~\cite{zrpc}, and some time will be devoted to explaining
how this approximate symmetry works.

\smallskip

Throughout the paper, $f$ stands for $f_{7/2}$ (except of course when
we speak of the $pf$ shell) and $r$, generically, for any or all of
the other subshells ($p_{1/2}\;p_{3/2}\;f_{5/2}$). Spaces of the type

\begin{equation}
\label{t}
f^{n-n_0} r^{n_0}+f^{n-n_0-1} r^{n_0+1}+\cdots+
f^{n-n_0-t} r^{n_0+t}
\end{equation}
represent possible truncations: $n_0$ is different from zero if more
than 8 neutrons (or protons) are present and when $t=n-n_0$ we have
the full space $(pf)^n$ for $A=40+n$.

We use: harmonic oscillator wave functions with $b=1.01 A^{1/6}$~fm;
bare electromagnetic factors in $M1$ transitions; effective charges of
1.5~$e$ for protons and 0.5~$e$ for neutrons in the electric
quadrupole transitions and moments.

Other definitions and conventions are introduced at the beginning of
the sections in which they are first needed.

\section{Level schemes and electromagnetic properties }
\label{sec:spectro}

The diagonalizations are done with the code {\sc
  antoine}~\cite{antoine}, a fast implementation of the Lanczos
algorithm in the $m$-scheme. Some details may be found in
ref.~\cite{cpz3}. The interaction KB3 is the same as in I, and it will
be revisited in section~\ref{sec:specfac}.  The $m$-scheme and maximal
$JT$ dimensions of the nuclei analyzed are given in
table~\ref{tab:dime}.

\subsection{Energy Levels in A=47}
\label{sec:el_47}

\noindent
$^{47}$Ca. Calculated and experimental levels are compared in
fig.~1. The ambiguities in the experimental spin assignments make it
difficult to pass judgement on the agreement, except that it seems
fair. 

\noindent
$^{47}$Sc (fig. 2). In this nucleus the experimental situation is
cleaner and the agreement striking. Notice the very nice
correspondence for the high spin states.

\noindent
$^{47}$Ti (fig. 3). The stable member of the $A=47$ isobaric multiplet
is the best known experimentally. We have plotted all the calculated
and measured levels of negative parity up to 4~MeV. From there up,
only the yrast or the ones with high spin, relevant to the discussion
below are included. There is an excellent correspondence between
theory and experiment; the ground state doublet, the triplet around
1.5 MeV and the two bunches of levels at 2.5 and 3.7 MeV. In the high
spin regime, Cameron and collaborators~\cite{cameron_47} have
identified two states at 6.366 and 8.005 MeV. They propose spins
$\frac{21}{2}^-$ and $\frac{27}{2}^-$ while our calculation favours
$\frac{19}{2}^-$ and $\frac{25}{2}^-$ in line with the assignments
made in Burrows' compilation~\cite{burrows_47}. The yrast sequence
will be discussed in detail in section~\ref{sec:collective}.

\noindent 

$^{47}$V and $^{47}$Cr (fig. 4). The almost degererate ground state
triplet of these nuclei is one of the hard nuts to crack in $pf$-shell
spectroscopy. In the MBZ~\cite{mbz} model, i.e. a single
$1f_{\frac{7}{2}}$ shell and the $^{42}$Sc two-body matrix elements as
effective interaction, the state $\frac{3}{2}^-$ appears at an
excitation energy of 1.2 MeV above the $\frac{5}{2}^-$-$\frac{7}{2}^-$
ground state doublet. The perturbative quasiconfiguration calculation
of~\cite{pozu2}(using the KB3 interaction) brought the $\frac{3}{2}^-$
state down to 0.5 MeV, but it takes the exact $pf$ shell
diagonalization to give the correct ordering $\frac{3}{2}^-$,
$\frac{5}{2}^-$, $\frac{7}{2}^-$.  A $\frac{3}{2}^-$ ground state is
consistent with $^{47}$V being a K=$\frac{3}{2}$ rotor. The
theoretical level scheme fits perfectly to the experiment except for a
couple of doublets that come out inverted.

\subsection{Electromagnetic Moments and Transitions in A=47}
\label{sec:et-mo_47}

In table~\ref{tab:muq47} we collect our predictions for the magnetic
dipole and electric quadrupole moments to compare with the availabe
experimental values. The agreement is very good, except for $Q$ in
$^{47}$Ca.  For $^{47}$V and $^{47}$Cr no experimental results are
known.

For $^{47}$Ca, there is an experimental upper bound to the probability
of the transition $3/2^-$~$\rightarrow$~$7/2^-$; $B(E2) <
2.8$~$e^2$~fm$^4$, that is compatible with the predicted value
(0.68~$e^2$~fm$^4$). Tables~\ref{tab:t_sc47},~\ref{tab:t_ti47}
and~\ref{tab:t_v47} display the E2 and M1 transition probabilities in
$^{47}$Sc, $^{47}$Ti and $^{47}$V. The data are of poor quality, but
in general compatible with our results, though there is a tendency for
quadrupole transitions between low-lying levels to be stronger than
our predictions indicate. This effect can be probably attributed to
mixing with the intruder excitations of the $^{40}$Ca core.

\subsection{Energy Levels in A=49}
\label{sec:el_49}

\noindent
$^{49}$Ca and $^{49}$Sc. These nuclei will be discussed in
section~\ref{sec:specfac}.

\noindent
$^{49}$Ti is the stable member of the $A=49$ isobaric multiplet. It is
interesting to compare fig.~\ref{fig:e_ti49} for this nucleus with
fig.~\ref{fig:e_sc47} for $^{47}$Sc, its cross-conjugate in the $f^n$
model, which would predict identical spectra. There are similarities
indeed, but there are also significant differences. The quality of the
agreement with experiment is high in both cases. Fig.~\ref{fig:e_ti49}
deserves a special comment: it was drawn using the information of the
1986 compilation of ref.~\cite{burrows_49}. In the 1995 version the
level immediately above 1.6 MeV that was given as $(5/2,7/2,9/2)^-$
becomes a doublet ${5/2}^-$-${9/2}^-$ leading to a one to one
corresponce with the calculations below 2 MeV, except for the new
${5/2^-}$ level---a possible intruder. However, the next two levels
that were taken to be either ${5/2}^-$ or ${7/2}^-$ in both cases, are
now given a tentative ${5/2}^-$ assignment, while the calculations
would obviously prefer ${7/2}^-$ \ldots

\noindent
$^{49}$V is our showpiece: the quality of the agreement in
figure~\ref{fig:e_v49} is simply amazing.

\noindent
$^{49}$Cr and $^{49}$Mn. The left panel of fig.~\ref{fig:e_crmn49}
indicates that up to the second 19/2$^-$ state there is a one to one
correspondence between the experimental levels and the theoretical
ones with an excellent agreement in energies. The data are consistent
with a $K=5/2$ rotor as we shall explain in
section~\ref{sec:collective}. The yrast states above 7 MeV in
$^{49}$Cr, taken from a recent experiment~\cite{cameron_49}, appear to
be 2~MeV above the calculated ones, a discrepancy well beyond the
typical deviations of our results. Furthermore, in the $f^n$ space
$^{47}$V and $^{49}$Cr are cross conjugate and have the same spectra:
It would be a real surprise to find a 25/2$^-$- 27/2$^-$ doublet 2~MeV
higher in Chromium than in Vanadium. There seem to be two ways out of
the conundrum.

\begin{itemize}
  
\item Assume the states are not yrast. The calculations have been
  pushed to include several states for each $J$, and some tentative
  correspondencess---based on energetics and decay properties--- are
  indicated in the figure by dot-dash lines. The identifications are
  hazardous, and we prefer the alternative explanation.

\item Assume the states are yrast. Then some gamma(s) may have been
  misplaced in the level scheme. The right panel of
  fig.~\ref{fig:e_crmn49} shows what would be the situation if the 
  $J=23/2^-$ level was  taken to decay to the first (lowest) $J=19/2^-$
  instead of the second (as assumed in~\cite{cameron_49}). Cross
  conjugation is now respected, and the agreement with the calculation
  becomes excellent. Therefore we shall adopt this interpretation of the
  spectrum in the discussion in section~\ref{sec:collective}.
\end{itemize}

\subsection{Electromagnetic Moments and Transitions in A=49}

The experimental information about magnetic moments is collected in
Table~\ref{tab:muq49}. In all cases the predictions agree with the
data, including the very recently measured moment of the 19/2$^-$
isomeric state in $^{49}$Cr. The only known $Q$ nicely agrees with the
calculated one.

In tables~\ref{tab:t_ti49},~\ref{tab:t_v49} and~\ref{tab:t_cr49} we
find the experimental information on $E2$ and $M1$ transitions for the
$A=49$ isotopes of Ti, V, and Cr. ( The data available for the other
members of the multiplet are too imprecise to compare with the
calculation.) The agreement is in general excellent, and in particular
for the $B(E2)$ values in $^{49}$V and $^{49}$Cr and the $B(M1)$
transition probabilities in $^{49}$Cr.

\smallskip

{}From this review of detailed properties it is possible to draw a
general conclusion: the calculations are very successful in describing
the data, the only systematic exceptions coming from the Ca isotopes:
The agreement in fig.~1 is probably the less satisfactory of all we
have shown, and in table~\ref{tab:muq47} the magnetic moment of
$^{47}$Ca is the only one that deviates significantly from the
measured ones. In I we had already had problems with the positioning
and B(E2) value for the $2^+$ state of $^{48}$Ca, and in
section~\ref{sec:specfac} we shall also encounter some discrepancies
of monopole origin in $^{49}$Ca. When corrected, the agreement with
experiment will no doubt improve, but it seems almost certain that a
full understanding of the Ca isotopes demands a closer look at the
influence of intruders.
  
\section{M1 and GT strength functions}
\label{sec:m1gt}

In this section we shall calculate strength functions following
Whitehead's prescription~\cite{white}. The procedure amounts to define
a ``sum rule state'' by acting with the transition operator we are
interested in ( M1 or GT here, single particle creation operator in
next section) and then use it as starting state for Lanczos iterations.
At the $n$-th iteration $n$ peaks are obtained that contain full
information on the $2n-1$ moments of the strength
distribution. The sum rule is the norm of the starting state. For more
detailed explanations and illustrations the reader is referred
to~\cite{cpz1,cpz2,bloom}.

For the M1 strength we write
\begin{equation}
  \label{M1}
  B(M1)= \left(\sqrt B_l \pm \sqrt B_s\right)^2,
\end{equation}
 where the $l$ and $s$ subscripts correspond to orbital and spin
 contributions respectively.

The Gamow-Teller (GT) strength is defined through
\begin{equation}
  B(GT)=\left(\frac{g_A}{g_V}\right)^2_{\text{eff}} \langle
  \bbox{\sigma\tau} \rangle^2, \hspace{0.5cm}
  \langle\bbox{\sigma\tau}\rangle =\frac{\langle f||\sum_k
    \bbox{\sigma}^k \bbox{t}^k_\pm ||i\rangle}{\sqrt{2J_i+1}},
\end{equation}
where the matrix element is reduced with respect to the spin operator
only (Racah convention~\cite{edmonds}), $\pm$ refers to $\beta^\pm$
decay, $\bbox{t}_\pm = (\bbox{\tau}_x \pm \text{i}\bbox{\tau}_y)/2$,
with $\bbox{t}_+ p = n$ and $(g_A/g_V)_{\text{eff}}$ is the effective
axial to vector ratio for GT decays,
\begin{equation}
  \left(\frac{g_A}{g_V}\right)_{\text{eff}} = 0.77
  \left(\frac{g_A}{g_V}\right)_{\text{bare}},
\end{equation}
with $(g_A/g_V)_{\text{bare}} = 1.2599(25)$~\cite{towner};

 for Fermi decays we have
    \begin{equation}
     B(F)=\langle\bbox{\tau}\rangle^2, \hspace{1cm}
   \langle\tau\rangle =\frac{\langle f ||\sum_k
      \bbox{t}^k_\pm || i\rangle}{\sqrt{2J_i+1}};
    \end{equation}

Half-lives, $T_{1/2}$, are found through

\begin{equation}
(f_A+f^\epsilon)\,T_{1/2}=\frac{6146\pm6}{(f_V/f_A)B(F)+B(GT)}.
\end{equation}
We follow ref.~\cite{wilki} in the calculation of the $f_A$ and $f_V$
integrals and ref.~\cite{bamby} for $f^\epsilon$. The experimental
energies are used.

\subsection{Scissors mode  in $^{47}$Ti}
\label{sec:ti47colec}

The existence of magnetic dipole orbital excitations in deformed
nuclei has been a topic of interest since the discovery of the
``scissors'' mode in heavy nuclei~\cite{richter}. It was suggested by
Zamick and Liu~\cite{zamick} that scissors-like excitations coud also
exist in lighter nuclei. In a study of $^{46}$Ti and
$^{48}$Ti~\cite{cpz3} it was found that the low lying 1$^+$ states of
these nuclei has indeed orbital character. In odd nuclei the strength
can split in up to three $J$ values. Since experiments can be done
only for stable targets, we have repeated the calculations for
$^{47}$Ti and $^{49}$Ti. For the latter, nothing very interesting
seems to happen, since quadrupole collectivity---an essential
ingredient in scissor-like behaviour---is practically absent. For the
former, quadrupole coherence is fairly strong, as we shall see in
section~\ref{sec:incipient}. Therefore we restrict ourselves to
showing in fig.~\ref{fig:ti47m1} the orbital, spin and total $M1$
strength functions of $^{47}$Ti.  Notice the very different structure
of the orbital and spin strengths.  The former is dominated by the
peaks at around 2 MeV while the latter is basically a resonace
centered at about 9 MeV with a much smaller peak at low energy. In the
total strength the two regions show up. The large spikes in the
orbital strength are the natural candidates to represent the scissors
mode. The angular momentum, excitation energy and strength of these
states are:

\begin{center}
\renewcommand{\arraystretch}{1.2}
\setlength{\tabcolsep}{0.3cm}
\begin{tabular}[h]{ccccc}
 $J$ & E (MeV) & Orbital & Spin & Total \\
 7/2$^-$ &  2.34 & 0.162 & 0.229 & 0.776 \\
 3/2$^-$ &  2.37 & 0.092 & 0.093 & 0.370
\end{tabular}
\end{center}

The values of the ratio $B_l/B_s$, 1.0 and 0.7 (using bare gyromagnetic
factors), strongly support the scissors interpretation of these
magnetic dipole excitations.

\subsection{Half-lives and Gamow-Teller strength functions}
\label{sec:strength_47}

We denote the total strength in the $\beta^+$ or $(n,p)$ channel by
$S_+$ and the strength in the $\beta^-$ or $(p,n)$ channel by $S_-$.
We express them in units of the Gamow-Teller sum rule so that they
satisfy $S_- - S_+ = 3(N-Z)$. From past experience we know that even
severely truncated calculations may give a sensible view of the
overall strength distributions, but miss the sum rule values by a
sizeable factor~\cite{a48,fierros}. To have a fuller picture we have
compared exact results with those of the most severe truncation
compatible with the $3(N-Z)$ sum rule, $t=0$ in the father and $t=1$ in
the daughters (notation defined in eq.~(\ref{t})) . The results are
compiled in table~\ref{tab:smas} incorporating the numbers we had for
$A=48$.  There is, in all cases, a strong reduction of the $S_+$
strength. The reduction factor is fairly constant for each isotopic
chain and all the values are close to 2. This is probably a good
estimate of the reduction of strength to be expected in the general
case due to $0\hbar \omega$ correlations.

The half-lives of the $A=47$ isobars are known up to $^{47}$Cr.  In
table~\ref{tab:half47} we compare our calculations (using the
effective value of $g_A$) with the experimental values. Agreement is
perfect for Ca, Sc and Cr and fair for V. For the proton richer nuclei
the half-lives are not known and we list our predictions alone. For
the $Z>N$ nuclei we also show the percentage of intensity that goes to
the analog state by a Fermi transition. Where it is experimentally
known ($^{47}$Cr) the agreement with the predicted value is very good.
In table~\ref{tab:half49} we proceed in the same way with the $A=49$
multiplet. All the calculated half-lives have been measured. The
agreement is close to perfect for all the short ones. Very long
half-lives usually mean that the decay window is very small and the
fraction of strength allotted to it is very
critical, which probably explains  the discrepancies in the Ca and V
cases.

Figure~\ref{fig:sc47beta} shows the strength function for the
processes $^{47}$Sc$(\beta^-){}^{47}$Ti and
$^{47}$Fe$(\beta^+){}^{47}$Mn. The spikes that come out of the Lanczos
strength function have been smoothed by gaussians of 500~keV full
width half maximum (FWHM) if they correspond to converged or
quasi-converged states and 1.3~MeV otherwise.  The Fe to Mn decay---
not studied experimentally yet---has a very large $Q_\beta$ window
($Q_{\text{EC}}=15.64$~MeV) that covers an important fraction of the
full strength.  On the contrary, only the small bump at around 0~MeV
in the strength function contributes to the $^{47}$Sc decay (here we
are lucky and the half-life comes out on the experimental spot).

Next, we show the strength functions corresponding to the---not yet
performed---reactions $^{47}$Ti$(p,n){}^{47}$V and
$^{47}$Ti$(n,p){}^{47}$Sc (figure~\ref{fig:ti47gt}). We have followed
different procedures to present the distributions of the $(p,n)$ and
$(n,p)$ processes. For the former, the individual peaks are broadened
by Gaussians whose width is taken to be equal to the typical
instrumental one for these reactions. For the latter, the spikes have
been replaced by gaussians with FWHM=1~MeV, and then we have summed up
the strength in 1~MeV bins. We show the original spikes as 
reference. The upper part of the figure contains the strength function
obtained in a $t=1$ calculation. It is similar in structure to the
full calculation (lower left part) though the resonance is shifted
down by some 2.5~MeV.  For the reactions $^{49}$Ti$(p,n){}^{49}$V and
$^{49}$Ti$(n,p){}^{49}$Sc, figure~\ref{fig:ti49gt} shows that the $t$=1
calculation gives a fair idea of the exact distribution, but the
Gamow-Teller resonance is again shifted down by some 2-3~MeV.

Comparing the exact $(p,n)$ profiles in figures~\ref{fig:ti47gt}
and~\ref{fig:ti49gt} we find that the GT resonance is definitely
broader in $^{47}$Ti than in $^{49}$Ti. The effect does not show in
the truncated calculations, which suggests that the extra broadening
should be associated to deformation---absent in $^{49}$Ti---but
significant in $^{47}$Ti as we shall see in
section~\ref{sec:collective}.

\section{Spectroscopic Factors in $^{49}$S\lowercase{c} and
  $^{49}$C\lowercase{a}. The KB3 interaction.}
\label{sec:specfac}

The monopole modifications that cure the defficiencies of the KB
matrix elements and transform them into the excellent KB3 interaction
can be characterized by a single prescription:

\smallskip

{\it make sure to have correct gaps and correct single particle
  properties in} $^{48}$Ca {\it and} $^{56}$Ni.

\smallskip

Our purpose in this section is to analyse in detail this prescription.

\subsection{Elementary monopole results}
\label{mono}
The reason to single out closed shells and single-particle and
single-hole states built on them ($cs\pm 1$ for short) is that we know
to a good approximation their eigenstates, whose energies are given in
terms of the few parameters that define the ``monopole'' Hamiltonian.
Writing $H=H_m+H_M$, as a sum of monopole($m$) and multipole($M$)
parts, and calling $n_r$ and $T_r$ the number and isospin operators
for orbit $r$ of degeneracy $D_r=2(2j_r+1)$, $V_{rstu}^{JT}$ the
two-body matrix elements and $\varepsilon_r$ the single particle
energies , we have ~\cite{FR69}:

\begin{mathletters}
\begin{eqnarray}
 &&H_{m} = \sum_r \varepsilon_r n_r + \label{monoa} \\
&&\sum_{r\leq s}(1+\delta_{rs})^{-1}[ a_{rs}\,n_r(n_s-\delta_{rs})
             +b_{rs}(T_r\cdot  T_s-\frac{3}{4}n \delta_{rs})],
\nonumber
\end{eqnarray}
\begin{equation}
a_{rs}=\frac{1}{4}(3V^1_{rs}+V^0_{rs}),\quad
b_{rs}=V^1_{rs}-V^0_{rs},
\label{monob}
\end{equation}
\end {mathletters}
where the ``centroids'' are:
\begin{equation}
V_{rs}^T=\frac{\sum_J V_{rsrs}^{JT}(2J+1)}{\sum_J (2J+1)}.
\label{centroid}
\end{equation}
The sums run over Pauli allowed values of $J$.
The important property of $H_m$ is that it reproduces the average
energies of the configurations to which a given state belongs. For
the $cs\pm 1$ set, there is only one state per configuration, and
therefore its energy is exactly given by $H_m$.

Calling $f\equiv f_{7/2}$, $p\equiv p_{3/2}$, $r\equiv
p_{3/2},p_{1/2},f_{5/2}$, and $r'\equiv p_{1/2},f_{5/2}$, from
eqs.~(\ref{monoa}) and~(\ref{monob}) we find the following estimates
for binding energies ($B_e$), single particle gaps ($\Delta$) and
excitation energies ($\epsilon$):

 \begin{equation}
   \label{e_48}
   B_e(^{48}\text{Ca})=8\varepsilon_f+28V^1_{ff}.
 \end{equation}
 \begin{equation}
   \label{e_56}
   B_e(^{56}\text{Ni})=16\varepsilon_f+120a_{ff}-6b_{ff}.
 \end{equation}
 \begin{eqnarray}
   \text{$\Delta$}(^{48}\text{Ca})&=&
 -2 B_e(^{48}\text{Ca})+B_e(^{49}\text{Ca})+
B_e(^{47}\text{Ca})
\nonumber\\
&=&\varepsilon_p-\varepsilon_f+8V_{fp}^1-7V_{ff}^1.
   \label{gap_48}
 \end{eqnarray}
 \begin{eqnarray}
  \text{$\Delta$}(^{56}\text{Ni})&=&
 -2 B_e(^{56}\text{Ni})+B_e(^{57}\text{Ni})+
B_e(^{55}\text{Ni})
\nonumber\\
&=&\varepsilon_p-\varepsilon_f+16a_{fp}-15a_{ff}+\frac{3}{4}b_{ff}.
   \label{gap_56}
 \end{eqnarray}
 \begin{equation}
   \label{e_49ca}
   \epsilon_{r'}(^{49}\text{Ca})=\varepsilon_{r'}-\varepsilon_p
+8(V^1_{fr'}-V^1_{fp}).
 \end{equation}
  \begin{equation}
   \label{e_49sc}
   \epsilon_{r}(^{49}\text{Sc})=\varepsilon_{r}-\varepsilon_f
+8(a_{fr}-a_{ff})-\frac{5}{2}(b_{fr}-b_{ff}).
 \end{equation}
 \begin{equation}
   \label{e_57ni}
   \epsilon_{r'}(^{57}\text{Ni})=\varepsilon_{r'}-\varepsilon_p
+16(a_{fr'}-a_{fp}).
\end{equation}
 
{\bf NOTE} In work on masses---to avoid minus signs---it is customary
to take $B_e>0$ for bound systems. Here we keep $B_e<0$, but reverse
the definition of $\Delta$ so as to conform with the usual one
($\Delta>0$ means the closed shell is more bound).

\smallskip

The expressions above are useful guides, but the $a_{rs}$ and $b_{rs}$
parameters must be chosen so that the {\em exact} diagonalizations
reproduce the binding and excitation energies. At the time the
monopole modifications to KB were proposed the task involved some
guessing, that can now be eliminated, and it is instructive to
reexamine critically the KB3 interaction, which was defined in three
steps~\cite{pasquini,pasquiniz,pozu1,pozu2}:
\begin{description}
\item[KB'] The $V_{fr}^T$ centroids.
\begin{equation}
\label{eq:kb'}
V_{fr}^T(\text{KB'})=V_{fr}^T(\text{KB})-(-)^T\,300 \text{ keV},
\end{equation}
\item [KB1] The $V_{ff}^T$ centroids ($V_{fr}^T$ centroids from KB')
\begin{equation}
  \begin{array}{l}
    V_{ff}^0(\text{KB1}) = V_{ff}^0(\text{KB})-350 \text{ keV},\\[2mm]
    V_{ff}^1(\text{KB1})= V_{ff}^1(\text{KB})-110 \text{ keV}.
  \end{array}
  \label{eq:kb1}
\end{equation}
\item [KB3] KB1 plus minor non-monopole changes.
\begin{equation}
  \begin{array}{l}
    W_{ffff}^{J0}(\text{KB3}) = W_{ffff}^{J0}(\text{KB1}) - 300
    \text{ keV for $J=1,3$};\\[2mm]
    W_{ffff}^{21}(\text{KB3}) = W_{ffff}^{21}(\text{KB1})-200 \text{
      keV}.
  \end{array}
  \label{eq:kb3}
\end{equation}

while the other matrix elements are modified so as to keep the
centroids~(\ref{eq:kb1}). These very mild changes
were made to improve the spectroscopy of some nuclei at the beginning
of the $pf$ shell. Their limited influence is discussed in I, and we
shall not bother with them in this paper.

\end{description}
   
To give an idea of the influence of the monopole changes:
Eqs.~(\ref{gap_48}) and~(\ref{gap_56}) yield

$\Delta$($^{48}$Ca)=2.06~MeV, and $\Delta$($^{56}$Ni)=3.42~MeV for KB,
 
$\Delta$($^{48}$Ca)=4.46~MeV, and $\Delta$($^{56}$Ni)=5.86~MeV for KB',
 
$\Delta$($^{48}$Ca)=5.22~MeV, and $\Delta$($^{56}$Ni)=8.57~MeV for KB3,

to be compared with experimental values

$\Delta$($^{48}$Ca)=4.81~MeV, and $\Delta$($^{56}$Ni)=6.30~MeV (exp).
 
If we turn to the exact calculations, we find that
$\Delta$($^{48}$Ca), for KB increases by some 600 keV, while for KB'
and KB3 it hardly moves!  For KB3 it actually goes down to 5.17~MeV.
For $^{56}$Ni, the situation is more difficult to assess because full
diagonalizations are not possible yet. Still, going to the $t=4$ level
reveals that for KB, instead of a closed shell we have a nice
rotational band dominated by 4p-4h configurations. At the same level
of truncation the KB' ground state remains normal but we are still far
from convergence, and at $t=6$ it produces in turn a rotational ground
state. KB3 yields $\Delta$($^{56}$Ni)=7.90~MeV, and extrapolation to
the exact result indicates that the closed shell will persist with the
gap remaining above the experimental one by about
1~MeV~\cite{fierros}.

\subsection{Spectra and spectroscopic factors}
\label{spectro} 
Let us now examine the spectrum of $^{49}$Ca. In fig.~\ref{fig:e_ca49}
we show the results for KB, KB' and KB3. The unmodified interaction
predicts 10 levels below 3.2~MeV, where two are observed. The
remarkable thing is that with the change in $V^1_{fr}$---involving a
single parameter---KB' produces a phenomenal improvement. The
remaining discrepancies are eliminated by one extra modification in
$V^1_{ff}$: The quality of agreement with experiment achieved by KB3
is excellent for the levels below 4.1~MeV, with two possible
exceptions:

The calculations predict two ${7/2}^-$ levels at 3.04 and 4.09~MeV
with very small spectroscopic factors for $(d,p)$ transfer, but still
sufficient to be observed. Both are strongly dominated by the $f^7p^2$
configuration, and very unlikely to move up by more than a few hundred
keV.  Now: there are possible experimental counterparts at 3.35 and
4.1~MeV, assigned as ${9/2}^+$ and ${7/2}^+$ respectively. {\em We think
there is room for revising these assignments, especially for the first
of the two states.}

\smallskip

For $^{56}$Ni we have not gone beyond the calculations
in~\cite{fierros}, which indicate that the single particle spectrum is
quite adequately described by KB3. As made clear by
eqs.~(\ref{e_49sc}) and~(\ref{e_57ni}), $^{49}$Sc should provide the
same information about centroids as $^{57}$Ni. Fig.~\ref{fig:e_sc49}
does not seem very encouraging in this respect: Although there is a
nice correspondence between theory and experiment for the first bunch
of levels around 4~MeV, it is impossible to read from the spectra
alone any clear message about single particle excitations, and hence
about monopole behaviour. The reason is that the states we are
interested in,

\begin{equation}
   \label{pivot}
  |r\rangle= a^{\dag}_r|^{48}\text{Ca gs}\rangle,
\end{equation}
are heavily fragmented. The amplitude of vector $|r\rangle$ in each of the
fragments is essentially the spectroscopic factor, defined as
\begin{equation}
    \label{eq:spec}
    S(j,t_z)= \frac{\langle J_f T_f T_{zf} || a^{\dag}_{jt_z} || J_i
      T_i T_{zi} \rangle^2}{2 J_f +1},
\end{equation}
where the matrix element is reduced in angular momentum only; $j$ and
$t_z$ refer to the spin and third isospin component of the stripped
nucleon. To calculate $S(j,t_z)$ we proceed as in
section~\ref{sec:m1gt}: use $|r\rangle$ as starting vector for a sequence of
Lanczos iterations.

The excitation energies of the starting vectors,
\mbox{$e_r=\langle r|H|r\rangle-\langle f|H|f\rangle$,}  (in MeV)
 \begin{equation}
   \label{er49sc}
   e_{p_{3/2}}=4.54 \quad e_{p_{1/2}}=5.99 
\quad e_{f_{5/2}}=5.76,
\end{equation}
are almost identical to the values obtained through
eq.~(\ref{e_49sc}):
 \begin{equation}
   \label{eps49sc}
   \epsilon_{p_{3/2}}=4.58 \quad\epsilon_{p_{1/2}}=5.99 
\quad\epsilon_{f_{5/2}}=5.66,
\end{equation}
a result readily explained by the weakness of the ground state
correlations in $^{48}$Ca. By the same token the sum rules for
$(2j+1)S(j,t_z)$ are very close to their theoretical maximum,
$(2j+1)$. It may be worth mentioning here that the the sum rule is
actually quenched by a factor of about 0.7 because of the deep
correlations that take us out of the model space. The problem is
discussed in detail in ref.~\cite{quench}, and shall be ignored here.

\smallskip

The spectroscopic factors for $f_{7/2}$ in fig.~\ref{fig:spec_f7},
show little fragmentation. For $p_{3/2}$, fig.~\ref{fig:spec_p3} seems
to indicate a discrepancy between theory---that produces substantial
fragmentation---and experiment, that falls quite short of the sum
rule, by detecting basically only two peaks. Note that the higher, at
around 11.5~MeV, corresponds to the IAS of the ground state of
$^{49}$Ca.  The discrepancy is explained when we consider
fig.~\ref{fig:spec_p1} for the $p_{1/2}$ strength: now the too
numerous experimental fragments abundantly exceed the sum rule. What
seems to be happening is that the method chosen to analyse the data
does not distinguish among the $L=1$ peaks those with $J=1/2$ from
those with $J=3/2$, unduly favouring the former.

 Note that the lowest
calculated state is a bit too high in fig.~\ref{fig:spec_p1}. (The
higher states are again isobaric analogues.)

\smallskip
  
The situation becomes truly satisfactory for the $f_{5/2}$ strength in
fig.~\ref{fig:spec_f5}. The four lowest theoretical peaks may demand a
slight downward shift but they have nearly perfect counterparts in
experiment, where a fifth state also shows---a probable intruder.
Higher up the agreement remains quite good, especially if we remember
that, at 60 iterations, the spikes beyond 7~MeV do not represent
converged eigenstates but doorways, subject to further fragmentation.
Note that it is the second of the two IAS levels slightly above 15~MeV
that carries most of the strength.

It is clear that in the $J=1/2$ and 3/2 spectra, the lowest state in
each is sufficiently dominant to be identified as {\em the} single
particle state, while for $J=5/2$, this object has been replaced by a
bunch of levels at around 5~MeV. In all cases the calculations are too
high by some 300 to 500 keV, indicating unperturbed positions in
eq.~(\ref{e_49sc}) too high by about this amount: A residual monopole
defect that should be corrected. However, it is worth noting that the
main strength (i.e., {\em the} single particle states, and the 5/2
multiplet) has already been pushed down by about 1~MeV. This is a
genuine dynamical effect---abundantly studied in the literature
under the name of ``particle-vibration coupling ''---which amounts
to stress that when a particle is added to a ``core'', it couples not
only to its ground state, but also to its excitations.

What we have shown in this section is a fully worked out realistic
example of how the displacement and fragmentation of the original
``doorway'' states, $|r\rangle$, takes place. In the language of
Landau's theory one would speak of bare particles becoming dressed
quasi-particles, and it is one of the merits of exact shell model
calculations to be able to illustrate in detail this subtle dressing
process.

\subsection{A critical assessment of KB3}
\label{sec:assess}

Our basic tenet is that once $H_m$ provides the correct unperturbed
energies, the residual $H_M$ takes good care of the mixing. It is
particularly well illustrated by the study of the spectroscopic
factors, where both the unperturbed positioning ($H_m$) and the
fragmentation mechanism ($H_M$) are crucial. Other properties ---in
particular collective ones---are, usually, not as sensitive to
monopole behaviour, and if they are, it may be more difficult to have
a clearcut picture of the relative influence of $H_m$ and $H_M$.  What
we are aiming at, is that the place to look for monopole problems and
cures is in the $cs\pm 1$ states: From what we have seen in this
section, good gaps and single particle properties around $^{48}$Ca are
sufficient to ensure that the interaction is good for the nuclei at
the beginning of the $pf$ shell.

The outstanding problems seem relatively minor: the gap in $^{56}$Ni
should be reduced by about 1~MeV, and the $\epsilon_r$ values in
eq.~(\ref{eps49sc}) should be reduced by some 300 to 500~keV. The gap
in $^{48}$Ca needs also a reduction of some 300~keV and some gentle
tampering with $\epsilon_r$ values in $^{49}$Ca may be warranted. The
necessary changes amount basically to making $a_{fr}-a_{ff}$ more
attactive by 50~keV or so.

There is a serious problem though, with the KB3 interaction: the
$V^T_{rr'}$ centroids were left untouched, because for the nuclei we
had been interested in, their influence was small and difficult
to detect. In a sense this was a blessing since it simplified the task
of doing the monopole corrections. However, to move beyond $^{56}$Ni
it is necessary to do the corrections, because---as was pointed out by
Brown and Ormand~\cite{ormand}---the single hole properties of KB3
around $^{79}$Zr are atrocious, and the reader is warned {\em not} to
rely on this interaction above A$\approx$60 .  We shall not go
into the problem here, and simply refer to a forthcoming
characterization of $H_m$,---in terms of very few parameters---valid
for the whole periodic table~\cite{mono}.

\subsection{Binding energies}
\label{sec:be}

In I it was noted that KB3 overbinds all A=48 nuclei by about 780 keV
and it was proposed to cure the problem by a monopole correction of
28.85$n(n-1)/2$~keV. Calculating Coulomb energies, as in I,
through the following expressions $(n=\pi+\nu,\pi$ = protons, $\nu$ =
neutrons)~\cite{pasquini}

\begin{equation}
  \label{eq:Hcoul}
  \begin{array}{cc}
    \multicolumn{2}{c}{\displaystyle{H_{\text{Coul}} =
        \frac{\pi(\pi-1)}{2} V_{\pi\pi} + \pi\nu V_{\pi\nu} + 7.279
        \pi \text{ MeV,}}} \\[3mm]
    \displaystyle{V_{\pi\pi} = 0.300\,(50) \text{ MeV,}} &
    \displaystyle{V_{\pi\nu} = -0.065\,(15)}
    \text{ MeV,}
  \end{array}
\end{equation}

we obtain binding energies relative to the $^{40}$Ca core as
listed in table~\ref{tab:be_47} for A=47 and in table~\ref{tab:be_49}
for A=49. (Here we use the convention that binding energies are
positive).  The agreement with the measured values is quite good, the
larger discrepancies coreesponding to the estimates based on
systematics~\cite{mass}.

\section{Collective properties.}
\label{sec:collective}

The collective model~\cite{Bor-Mot2} provides a framework for the
study of the shape oscillations to which nuclei are subjected under
one form or another. In general, the coupling between different modes
precludes the sort of general predictions that are possible in the
extreme cases, and we shall start by examining $^{47}$Ti, which is
definitely collective, but neither a rotor nor a vibrator. Then we
move on to $^{49}$Cr-$^{49}$Mn and $^{47}$V-$^{47}$Cr, for which the
strong coupling limit should apply.  Following I, a good rotor will be
characterized by a $J(J+1)$ energy sequence {\em and} by a constant
intrinsic quadrupole moment $Q_0$ for all members of the band.  In
addition to being a constant, we expect $Q_0$ to be the same when
extracted from the spectroscopic quadrupole moment through

\begin{equation}
\label{eq:q0qsp}
Q_0=\frac{(J+1)\,(2J+3)}{3K^2-J(J+1)}\,Q_{spec}(J), \hspace{0.5cm}
  \text{for $K\neq 1$}
\end{equation}
or from the BE2 transitions through the rotational model prescription
  (for $K \neq  \frac{1}{2},1$)
\begin{equation}
\label{eq:q0be2}
B(E2,J\;\rightarrow\;J-2)=\frac{5}{16\pi}\,e^2|\langle JK20|
J-2,K\rangle|^2\, Q_0^2. 
\end{equation}

For even-even nuclei this is about as far as we can go in deciding
whether we are faced with a good rotor or not. When a particle is
added or removed, the collective model description of its coupling to
the rotor leads to some very precise predictions that make the
comparison with microscopic calculations more stringent. In
section~\ref{sec:mirrors} we are going to see that the low spin states
in the $^{49}$Cr-$^{49}$Mn and $^{47}$V-$^{47}$Cr mirror pairs follow
these predictions quite well. Then, a change of regime takes place and
the calculations remain in agreement with experiment but the strong
coupling limit of the collective model ceases to be valid.

By what should we replace it?
 
{}From the studies of $^{48,50}$Cr~\cite{cr48,cr50}, we know how the
change of regime---associated with backbend--- takes place, and there
is a framework---quasi-SU(3)~\cite{zrpc}---that describes yrast
behaviour before and after backbend. We shall devote some time to it
in section~\ref{sec:quasi}.

\subsection{Incipient collectivity in $^{47}$Ti.}
\label{sec:incipient}

Quadrupole collectivity is not confined to ``good rotors''. There are
other species that include ``vibrators'', ``$\gamma$-soft nuclei'' and
the like. $^{48}$Ti was found in I to be one of those nuclei that are
definitely collective, but definitely not rotors. Its neighbour
$^{47}$Ti belongs to the same category: In figure~\ref{fig:ti47rot} we
have plotted the experimental excitation energies against angular
momentum (open circles). A $J(J+1)$ law (open triangles) was fitted to
these points. The rms deviation is 313 keV and the static moment of
inertia ${\cal J}^{(1)} = 11$~MeV$^{-1}$.  The largest distortions
take place at spins $\frac{7}{2}^-$ and $\frac{9}{2}^-$. Clearly,
there is a rotational flavour, but it is not fully convincing. If now
we turn to the quadrupole moments, extracting $Q_0$ for the yrast
states from eqs.~(\ref{eq:q0qsp}) and~(\ref{eq:q0be2}) we find the
results shown in table~\ref{tab:q0_ti47}.  The $Q_0$ values coming
from the spectroscopic moments are quite erratic, while those obtained
from the $B(E2)$'s are closer to constancy at around $Q_0 =
60$~e~fm$^2$, similar to the number found in $^{48}$Ti.  Therefore,
though we do not have a good rotor, the structure of the yrast band of
$^{47}$Ti suggests the existence of an incipient prolate intrinsic
state.

Are we in the presence of a ``new form of collectivity''?

In attempting to give an answer, we may remember that Bohr's
collective Hamiltonian was precisely  designed to cope with such
intermediate coupling situations (see appendix 6B of~\cite{Bor-Mot2}).
Unfortunately, they demand the specification of potential energy
surfaces and inertial parameters that can be (meaningfully) determined
only in terms of ``some'' underlying microscopic Hamiltonian. 

Now: we have a reliable Hamiltonian that has proven capable
of describing whatever form of collectivity present in a given
nucleus. It also happens that the collectivity seems to be
predominantly quadrupole. Therefore, before (or perhaps, instead of)
answering the question above we should do well to examine the
following one(s):

What is the collective part of the Hamiltonian?
How does it work?

Section~\ref{sec:quasi} will be devoted to this problem.
     
\subsection{Particle-rotor coupling in $^{49}$Cr-$^{49}$Mn and
  $^{47}$V-$^{47}$Cr}
\label{sec:mirrors}

Since $^{48}$Cr seems to be a reasonably good rotor at low spins, we
expect the two mirror pairs obtained by adding and removing a particle
from it to be reasonably well described by the strong coupling limit
particle plus rotor model. In their simplest form, the predictions would
be the following:

\begin{description}

\item[Energies] The bands should have the same moment of the inertia
  as the underlying rotor. Coriolis decoupling is expected to be
  appreciable only in $K=1/2$ bands (not our case)

\item [Deformations] The quadrupole moments should be the same as
  those of the underlying rotor.

\item [Magnetic moments] The contribution of the extra particle (or
  hole) is now crucial, and can largely exceed that of the rotor. The
  precise prediction will be discussed in section~\ref{sec:gyro}.
   
\end{description}

At first we restrict ourselves to presenting evidence on these three
items. Conclusions will be drawn at the end of the section.

\subsubsection{Energetics and quadrupole properties}

\label{sec:enqua}

The yrast levels in figures~\ref{fig:e_crv47} and~\ref{fig:e_crmn49}
have been fitted to a $J(J+1)$ law, with the results shown in
fig.~\ref{fig:fit_47-49}. For $^{47}$V the agreement is quite good
for all spins , using a static moment of inertia, ${\cal J}^{(1)} =
12$~MeV$^{-1}$. For $^{49}$Cr only the states with $J \leq 17/2$ where
included in the fit, yielding, ${\cal J}^{(1)} = 8$~MeV$^{-1}$. The
energies of the states with $J \geq 19/2$ seem to behave as a strongly
decoupled rotational band of high $K$, but the quadrupole moments and
transitions do not support this interpretation.

The results of applying eqs.~(\ref{eq:q0qsp}) and~(\ref{eq:q0be2}) to
the $K=3/2$ and $K=5/2$ bands in the $^{47}$V-$^{47}$Cr and
$^{49}$Cr-$^{49}$Mn pairs respectively are given in
tables~\ref{tab:q0_47} and~\ref{tab:q0_49} . For the lowest states the
values are close to the ones obtained for $^{48}$Cr in ref.~\cite{a48}
($Q_0 (2^+) = 103\ e$~fm$^2$), which is indeed what we expect in the
particle plus rotor model, but the situation is somewhat different for
the two pairs:

\begin{itemize}

\item $^{47}$V-$^{47}$Cr. For $Q_0$ extracted from $Q_s$ there is 
 a curious staggering effect for the
  lower spins that does not show for the $BE2$-extracted numbers.
 The anomaly at $J=17/2$ is due to the presence of an almost
  degenerate state of this spin (refer to fig~\ref{fig:e_crv47}).
  Beyond this spin, $Q_0$ decreases but not drastically.
 In view of the good $J(J+1)$ behaviour for all spins,
  one may be tempted to conclude that there is no significant change
  of regime along the yrast line. The study of the gyromagnetic
  factors will tell us soon that this is not so.

\item $^{49}$Cr-$^{49}$Mn. Here the situation is simpler, and similar
  to that of $^{48}$Cr: A plot of $J\,vs\,E_{\gamma}$ would show a
  backbend at $J=17/2$. Beyond this spin the $f_{7/2}^n$ configuration
  becomes strongly dominant and $Q_0$ behaves erratically, and there
  are indications---to be confirmed by the gyromagnetic factors---that
   the change of regime starts at $J=13/2$, i.e., before the
  backbend.

\end{itemize}

 

\subsubsection{Gyromagnetic factors}
\label{sec:gyro}

A very interesting test of the validity of the collective model 
comes from the magnetic moments, for which the predicted contributions
of the particle (or hole) and the rotor are of similar magnitude.

For the gyromagnetic factors of a $K$-band we have:

\begin{equation}
  \label{eq:gr_pr}
   g(J)= g_R + (g_K-g_R) \frac{K^2}{J(J+1)}, \hspace{0.5cm} \text{for 
   $K\neq \frac{1}{2}$};
\end{equation}

where $g_R$ is the rotor gyromagnetic factor and $g_K$ is defined by

\begin{equation}
  \label{eq:gK}
  g_K K = \langle \Phi_K | g_l \bbox{l}_3 +
  g_s \bbox{s}_3|\Phi_K\rangle.
\end{equation}

 $|\Phi_K\rangle$ is the intrinsic wave function of the
particle, corresponding to a Nilsson orbit of  quantum numbers
$[Nn_zm_l]K$. Using the asymptotic wavefunctions we find

\begin{equation}
  \label{eq:gk_nilson}
   g_K = g_s + (g_l-g_s)\frac{m_l}{K}
\end{equation}

To compare the shell model results with eq.~(\ref{eq:gr_pr}) we have
to extract $g_R$ and $g_K$ from our wave functions. As a first step,
let us decompose the gyromagnetic factors as a sum of isoscalar
($g_0$) and isovector ($g_1$) contributions. Figure~\ref{fig:g0} shows
that $g_0$ is approximately constant for all the exact eigenvectors in
the mirror pairs and close to the rotor value ($^{48}$Cr). Therefore
we have $g_0\approx g_R $. The scale of the figure very much emphasizes
the discrepancies. For all practical purposes they can, and will, be
neglected.  For $^{48}$Cr the identification is trivial, since $ g_R $
is pure isoscalar.  For the other nuclei, the isoscalar contribution
to the collective prediction in eq.~(\ref{eq:gr_pr}) is

\begin{equation}
  \label{eq:g0_rot}
  g_0 = g_R + \left(\frac{g^\pi_K+g^\nu_K}{2} -
    g_R \right) \frac{K^2}{J(J+1)}.
\end{equation}

For  $g_0 \approx g_R$ to hold, the second term in the right hand side
must vanish, and then:

\begin{equation}
  \label{eq:gkgr}
   g^\pi_K-g_R \approx -(g^\nu_K-g_R).
\end{equation}

If we accept eq.~(\ref{eq:gkgr}) as a strict equality we have:

\begin{equation}
  \label{eq:g0g1}
  \begin{array}{l}
    \displaystyle{g_0 = g_R}, \\[2mm]
    \displaystyle{g_1 = (g^\pi_K -g_R)
      \frac{K^2}{J(J+1)}}.
  \end{array}
\end{equation}

We set $g_R = 0.535\ \mu_N$, the mean value of the gyromagnetic
factors of $^{48}$Cr.  To obtain $g_K-g_r$ we fit $g_1$ computed for
every spin using~(\ref{eq:g0g1}). We assume $K=3/2$ for
$^{47}$V-$^{47}$Cr and $K=5/2$ for $^{49}$Cr-$^{49}$Mn. The fit is
restricted to states with $J<19/2$ for $A=47$ nuclei and $J<15/2$ for
the $A=49$ nuclei. The values obtained for $(g^\pi_K - g_R)$ are

\begin{equation}
  \begin{array}{ll}
    g^\pi_{3/2}-g_R = 1.561\ \mu_N & \text{for }
    {^{47}\text{V}},\\  
    g^\pi_{5/2}-g_R = 1.137\ \mu_N & \text{for }
    {^{49}\text{Mn}},\\  
  \end{array}
\end{equation}

With $g_R = 0.535\ \mu_N$, and the expression~(\ref{eq:gkgr}) we
obtain for $g_K$ the numbers to the left in the rhs of
eq.~(\ref{eq:gk_smnil}).  For comparison, we have written to the right
the values predicted by eq.~(\ref{eq:gk_nilson}) for asymptotic
Nilsson wave functions [321]3/2 for $A=47$ nuclei and [312]5/2 for
$A=49$ using effective gyromagnetic factors for the proton and
neutron~\cite{Bor-Mot2}

\begin{equation}
  \label{eq:gk_smnil}
  \begin{array}{llll}
    g^\pi_{3/2} = \phantom{-}2.096\ & vs.\ &\phantom{-}2.130\  \mu_N
 & \text{for }    {^{47}\text{V}},\\ 
    g^\nu_{3/2} = -1.026\ & vs.\ & -1.023\  \mu_N
 & \text{for }    {^{47}\text{Cr}},\\ 
    g^\pi_{5/2} = \phantom{-}1.672\ & vs.\ &\phantom{-}1.718\  \mu_N
 & \text{for }    {^{49}\text{Mn}},\\
    g^\nu_{5/2} = -0.602\ & vs\ &  -0.654\  \mu_N
 & \text{for }    {^{49}\text{Cr}}. 
  \end{array}
\end{equation}

Clearly the agreement between both set of values is excellent. The
numbers to the right can be used to test eq.~(\ref{eq:gkgr}).  With
the set for $^{47}$V and $^{47}$Cr we find $g_R = 0.554\ \mu_N$,
while the one for $^{49}$Cr and $^{49}$Mn yields $g_R = 0.532\ \mu_N$,
which compare well with our adopted, $g_R = 0.535\ \mu_N$.

In figure~\ref{fig:47-49gr} we plot the gyromagnetic factors as a
function of spin. The dashed lines are the predictions of the particle
plus rotor model using the fitted values for $g_K$. The model gives a
very adequate average description of the exact results at low spin.
Deviations become important at $J=19/2$ for $^{47}$V-$^{47}$Cr and
$J=15/2$ for $^{49}$Cr-$^{49}$Mn, very precisely at the spins where a
change of regime is detected in tables~\ref{tab:q0_47}
and~\ref{tab:q0_49}.

\medskip

The conclusion is that the collective model works quite well in a
region where---until recently---it was not supposed to work. It can be
viewed as a classical background on which quantum deviations show as
staggering effects associated to Coriolis coupling. There is also a
discrepancy that---at first sight---seems perplexing: the odd nuclei
have a better $J(J+1)$ behaviour than the underlying rotor, and
different moments of inertia. Even more surprising: there is no
backbend in the $^{47}$V-$^{47}$Cr pair, though there is clearly a
breakdown of the strong coupling limit along the yrast band.

We do not have a detailed explanation for this behaviour, but a good
idea of its origin. In ref.~\cite{zrpc} it was shown that the
structure of the wavefunctions depends almost entirely on the
interplay between monopole and quadrupole forces, while the
backbending pattern is due to other parts of the interaction, and can
be treated in first order perturbation theory. (See figs.~2-3
of~\cite{zrpc}). The same is true for the moments of inertia (see
fig.~5 of ref.~\cite{cr48}). Therefore, the energetics may depend on
``details'' but the wavefunctions do not. The ``details'' can be very
important (pairing, for instance), and lead to secular effects, as the
observed correlations between moments of inertia and deformation.
Still, the monopole plus quadrupole contributions to the Hamiltonian
have special status.

The change of yrast regime at some critical spin is a very general
phenomenon and one would like to have a general answer to the
question: What happens at backbend? Our calculations certainly show
that the change in regime occurs at yrast energies where two levels of
the same spin are close together. This is in line with the
traditional interpretation in terms of ``band'' crossing, provided we
can associate the levels beyond backbend to a band. A preliminary
attempt in this direction was made for $^{50}$Cr in  ref.~\cite{cr50}.
The subject is certainly interesting, but we shall not pursue it here.

\subsection{Quasi-SU(3)}
\label{sec:quasi}

The collective model is a general framework to decide whether a given
nucleus is rotating or not. It says nothing concerning either the
onset or breakdown of rotational behaviour. Though it has been
suspected for long that Elliott's quadrupole force~\cite{elliott} is
the main agent responsible for stable deformations, with the advent of
successful phenomenological potentials of Skyrme and Gogny
type---capable of mean-field descriptions of bulk properties,
including deformation---the question went into a limbo. Recent studies
however, establish clearly that a quadrupole force of Elliot's type is
indeed massively present in the realistic forces~\cite{mdz}, and that
the rotational and backbending features in $^{48}$Cr originate in its
interplay with the monopole field~\cite{zrpc}. The crux of the matter
is that the latter may break Elliott's exact SU(3) scheme in such a
way that it is replaced by an approximate quasi-SU(3) symmetry. For
weak enough single particle splittings, rotational motion is almost as
perfect in the approximate scheme as in the exact one. When the
splittings are increased the backbending change of regime occurs.

The purpose of this section is to present some new results concerning
quasi-SU(3) and to illustrate its use in our region.

Let us start by considering fig.~\ref{fig:q-su3}. In the left panel we
have the spectra obtained by diagonalizing the operator
$2q_{20}=2z^2-x^2-y^2$ in the one particle space of the $pf$ shell. Two
particles can go into each level (only positive projections are
shown). By filling them orderly we obtain the intrinsic states for the
lowest SU(3) representations: $(\lambda ,0)$ if all states are occupied
up to a given level and $(\lambda ,\mu)$ otherwise. For instance:
putting two neutrons and two protons in the $K=1/2$ level leads to the
(12,0) representation. For four neutrons and four protons, the filling
is not complete and we have the (triaxial) (16,4) representation for
which we expect a low lying $\gamma$ band.

In the right panel of the figure we have the spectrum for the same
diagonalization but now in the restricted space of the $f_{7/2}$ and
$p_{3/2}$ orbits (the lower $\Delta j=2$ sequence, which we call $fp$).
The result is not exact, but a very good approximation. The idea is
that the orbits that come lowest after spin orbit splitting in a major
shell, form sequences $j$= 1/2, 5/2, 9/2\ldots or 3/2, 7/2, 11/2\ldots,
whose behaviour must be close to that of the sequences $l$= 0, 2,
4\ldots or 1, 3, 5\ldots that span the one particle representations
of SU(3). Quasi-SU(3) amounts to start from SU(3) and make the replacements

\[ l\longrightarrow j=l+1/2\quad m\longrightarrow m+1/2\times
\text{sign}(m).\]

The correspondence is one-to-one and respects SU(3) operator
relationships, {\em except} for $m=0$, where it breaks down.
Therefore, the symmetry cannot be exact, but this turns out to be an
asset rather than a liability, because the new coupling scheme can
account quite well for a variety of experimental facts. The way to
proceed is simply to reason with the right panel of
fig.~\ref{fig:q-su3} as we would do with the left one. Then, both the
four and eight particle ``representations'' for $T=0$ will be axial,
while the ten particle $T=1$ ones would be triaxial. A first positive
indication is the absence of a $\gamma$ band in $^{48}$Cr (its
counterpart in the $sd$ shell, $^{24}$Mg, is triaxial). For $^{50}$Cr
we would expect a $\gamma$ band. Experiment and calculations give no
clear answer in this case and in ref.~\cite{cr50} it was promised to
return to the problem in this paper. Before we go into it, it is worth
gaining some further insight into the meaning of quasi-SU(3).

\smallskip

So far we have introduced quasi-SU(3) following the arguments
of~\cite{zrpc}, which rest on the fact the two lowest $\Delta j=2$
sequence of orbits, separated from the rest by the spin-orbit
splitting, are sufficient to ensure quadrupole coherence. It does not
mean that the higher orbits can simply be neglected. To study their
influence we start by noting that all indications point to the
validity of a description in which quadrupole and monopole terms are
clearly dominant~\cite{cr48,zrpc}. Therefore, we want to diagonalize
the schematic Hamiltonian

\begin{equation}
  \label{hmq}
  H_{mq}=\hbar \omega \left(\sum \epsilon_in_i
  -\frac{8\kappa}{(p+3/2)^4}q\cdot q\right), 
\end{equation}
 
where we we have borrowed from~\cite{mdz} the normalized form of the
quadrupole force that emerges naturally when it is extracted from a
realistic interaction. Here we use $q=(4\pi /5)^{1/2}r^2Y^2$, $r$ is
the dimensionless coordinate, $p$ the principal quantum number. Since
we are interested in situations of permanent deformation, $q_{20}$ is
expected to be a good approximate quantum number. Therefore, we could
obtain the intrinsic state by linearizing $H$, which amounts to a mean
field calculation. (Note here that we want a Hartree, {\em not} a
Hartree-Fock variation, so as to guarantee the exact SU(3) solution
for vanishing single particle splittings.) The operation amounts to
replacing $q\cdot q$ by $q_{20}q_{20}$, and demands some care since
$q_{20}$ is a sum of neutron and proton contributions
$q_{20}=q_{20}^{\nu}+q_{20}^{\pi}$.  The correct linearization for the
neutron operators, say, is then

\[q_{20}\, q_{20}\longrightarrow
 q_{20}^{\nu}\langle q_{20}^{\nu}+2q_{20}^{\pi}\rangle
 \approx \frac{3}{2}q_{20}^{\nu}\langle q_{20}\rangle,\] 

where we have assumed $\langle q_{20}^{\nu}\rangle\approx
\langle q_{20}^{\pi}\rangle$. Therefore we are left with

\begin{equation}
  \label{hmql}
  H_{mq0}=\hbar \omega \left( \varepsilon H_{sp}
  -\frac{3\kappa}{(p+3/2)^4} \langle 2q_{20}\rangle 2q_{20}\right),   
\end{equation}

which is a Nilsson problem with the coefficient of $2q_{20}$ under a
new guise. In the usual formulation it is taken to be one third of the
deformation parameter $\delta$:

\begin{equation}
  \label{beta}
  \frac{\delta}{3}=\frac{1}{4}\frac{\langle 2q_{20}\rangle}{\langle
  r^2\rangle}= 
  \frac{\langle 2q_{20}\rangle}{(p+3/2)^4}.
\end{equation}

By equating with the coefficient of $2q_{20}$ in eq.~(\ref{hmql}) we
find
\[
\kappa =\frac{4}{12}= 0.33,
\]
which can be interpreted as a ``derivation'' of the value of the
quadrupole coupling constant. Of course, we would like it to agree
with the value extracted from a realistic interaction. In fact, it
comes quite close to it. From ref.~\cite{mdz} we know that the bare
$\kappa$ is 0.22, and that it is boosted by 30\% through core
polarization renormalizations. If only the quadrupole force is kept in
a schematic calculation a further boost of 15\% is necessary to
account for the effect of the neglected contributions. We are left
with $\kappa= 0.33$. The agreement seems too good to be true, but the
reader is referred to~\cite{mdz} to check that we have not cheated.

Nilsson diagrams are shown in fig.~\ref{fig:nilsson}. The right panel
corresponds to the usual representation in which the levels are shown
as a function of the deformation parameter $\delta$. In
eq.~(\ref{hmql}) we have set $\varepsilon=1$ and $\hbar \omega
H_{sp}$= the single particle spectrum as given in $^{41}$Ca (basically
equidistant single particle orbits $f_{7/2}$, $p_{3/2}$, $p_{1/2}$,
$f_{5/2}$ with a splitting of 2~MeV). In the figure the centroid of the
spectrum is made to vanish.

In the left panel we have turned the representation around: since we
are interested in rotors, we start from perfect ones (SU(3)) and study
what happens under the influence of an increasing single particle
splitting. Here we have used $\langle 2q_{20}\rangle\approx 32-36$,
obtained for 4 neutrons and 4 protons filling the lowest orbits in
either side of fig.~\ref{fig:q-su3}. In principle this number should
be obtained self consistently, but we shall see that it varies little
as a function of $\varepsilon$. For $p=3$, and $\hbar \omega \approx
10$~MeV, this means in round numbers $\delta=0.25$.

The figures suggest that quasi-SU(3) operates in full at $\varepsilon
\approx 0.8$ where the four lowest orbits are in the same sequence as
the right side of fig.~\ref{fig:q-su3} (Remember here that the real
situation corresponds to $\varepsilon \approx 1.0$). The agreement
even extends to the next group, although now there is an intruder
([310]1/2 orbit). The suggestion is confirmed by an analysis of the
wavefunctions: For the lowest two orbits, the overlaps between the
pure quasi-SU(3) wavefunctions calculated in the restricted $fp$ space
and the ones in the full $pf$ shell exceeds 0.95 {\em throughout the
  interval} \mbox{$0.5< \varepsilon <1$}. More interesting still: the
contributions to the quadrupole moments from these two orbits vary
very little, and remain close to the values obtained at
$\varepsilon=0$ (i.e., from fig.~\ref{fig:q-su3}). To fix ideas: for
$^{48}$Cr we would have $Q_0 \approx 2\langle q_{20}\rangle
A^{1/3}\equiv 116 e$~fm$^2$, a quite useful estimate of the exact
$Q_0\approx 100 e$~fm$^2$.

These remarks readily explain why calculations in the restricted
$(fp)^n$ spaces account remarkably well for the results in the full
major shell ($(pf)^n$).

\smallskip

Now we can understand from the diagrams why $^{48}$Cr
is {\em not} triaxial, and why the expected $\gamma$ band in $^{50}$Cr
fails to materialize: calculations in the $(fp)^{10}$ space indicate
that its presence depends on the near degeneracy of
the [312]5/2 and [321]1/2 orbits, which is broken because the
effective $\delta$ is likely to be closer to 0.2 than 0.3. As a
consolation we are left with a weaker prediction: in $A=49$, there
should be a low lying $K=1/2$ band, and indeed there is a $J=1/2^-,
3/2^-$ doublet below 2 MeV in fig~\ref{fig:e_crmn49}, which
calculations in the $(fp)^{9}$ space predict unambiguously to be the
lowest states of a fairly good rotational sequence, that persists well
after the degeneracy between the $K=1/2$ and 5/2 levels is broken.

\smallskip

It is seen that the description of rotational features can proceed in
three steps.

\begin{description}

\item[quasi-SU(3)] No calculations are done. We simply use
fig.~\ref{fig:q-su3} to estimate $Q_0$ and find hints about low lying
bands beyond the ground state one.

\item[schematic diagonalization] Diagonalizations of the quadrupole
  force in the presence of a single particle (or more generally
  monopole) field are made in the $\Delta j=2$ spaces to check the
  hints from the previous item. The simplicity of the problem will
  certainly point to efficient computational strategies that will make
  it possible to enlarge the spaces and account for the full
  interaction perturbatively.

\item [full diagonalization] In principle they are necessary for very
  accurate detailed descriptions. In practice they will be seldom
  feasible. However the experience gained in the few cases they are,
  will be of great help in checking the reliability of the methods
  emerging from the preceeding steps.

\end{description}

We close by insisting on the fact that the microscopic realistic
collective Hamiltonian is to a very good first approximation the monopole
plus quadrupole $H_{mq}$. SU(3) and quasi-SU(3) can be viewed as the
geometric equivalents of the strong coupling limit of the collective
model, but there are other regimes and one should take seriously the
task of diagonalizing the schematic Hamiltonian {\em exactly}.
                                                                      
\section{Conclusions}
\label{sec:conclu}

The nuclei around $^{48}$Cr have two special characteristics: they
seem to be the lightest in which collective features typical of the
medium and heavy regions appear, and they are the ones in which the
largest exact shell model calculations are possible at the moment. The
combination is a fortunate one, and we have tried to make the most of
it.

The results in section~\ref{sec:spectro} strongly support the
contention that realistic interactions with minimal monopole
modifications are capable of describing nuclear properties with great
accuracy. In many cases of dubious or ambiguous experimental
assignments we have proposed alternative ones. They could be
interpreted as ``predictions'' that could provide checks on our
results. Further checks could come through measures proposed in
section~\ref{sec:m1gt}.

In section~\ref{sec:specfac} we went to some length to explain what is
involved in the monopole modifications to the interaction. The
calculation of spectroscopic factors---necessary to disentangle the
monopole centroids from the raw data---is of interest in exploring in
detail particle-vibration coupling.

Finally, in section~\ref{sec:collective}, we dealt with particle-rotor
coupling. The aim was to understand what the calculations were
saying. First in terms of the collective model, and then in terms of
its basic microscopic counterpart, the quadrupole plus monopole
Hamiltonian, which exhibits an approximate symmetry---quasi-SU(3).
It may come as a surprise that things as simple as Nilsson diagrams in
their original form (oscillator basis in one major shell) could be
further simplified through quasi-SU(3), and then come
so close to describing what realistic interactions are doing in exact
diagonalizations. But then, the surprise is a pleasant one.

\bigskip

This work has been partially supported by the DGICyT, Spain under
grants PB93-263 and PB91--0006, and by the IN2P3-(France) CICYT
(Spain) agreements. A. P. Z. is Iberdrola Visiting Professor at the
Universidad Aut\'onoma de Madrid.

\onecolumn

\begin{figure}
  \begin{center}
    \leavevmode
    \epsfysize=10cm
    \epsffile{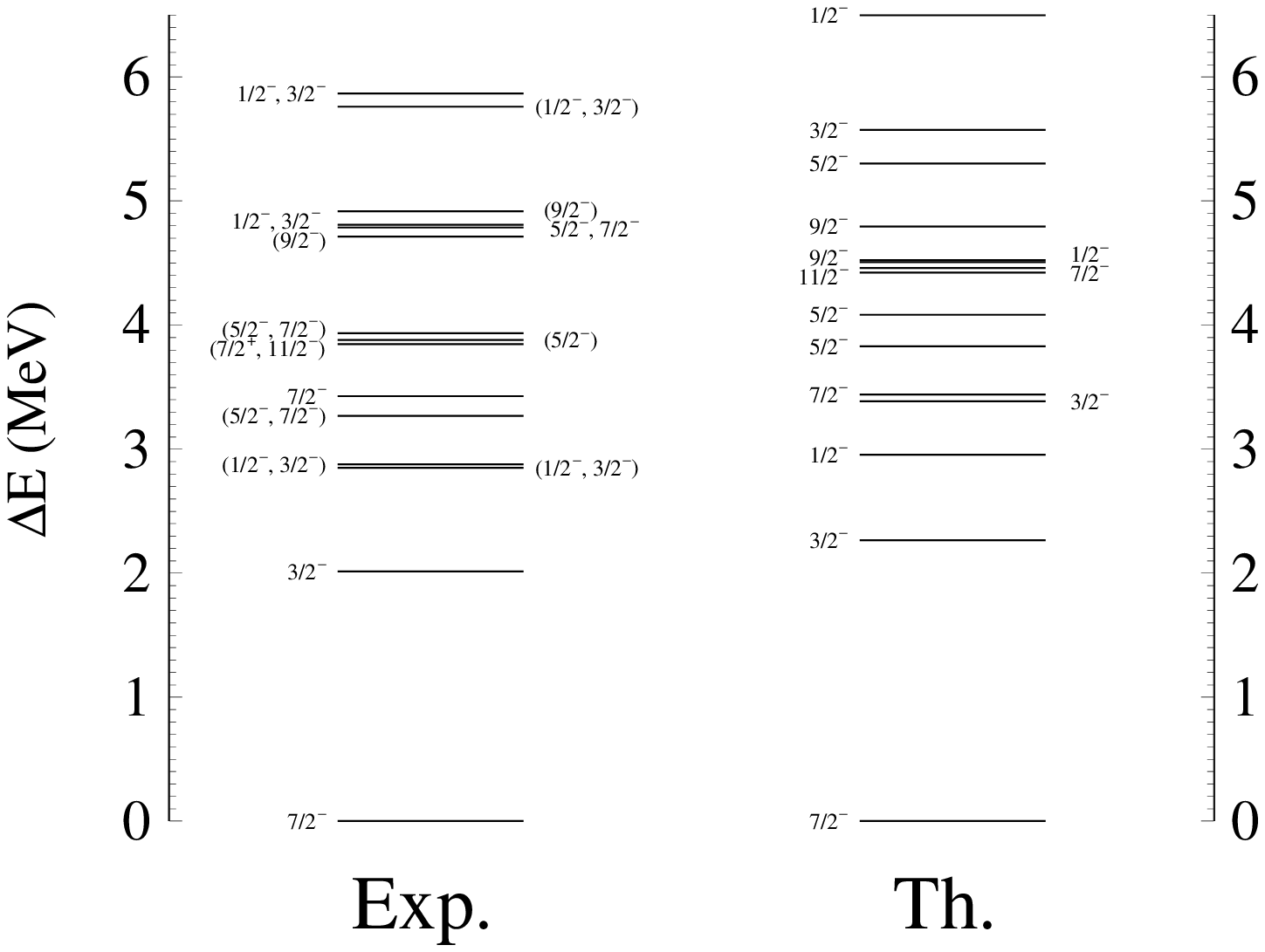}
    \caption{Theoretical and experimental energy levels of $^{47}$Ca.}
    \label{fig:e_ca47}
  \end{center}
\end{figure}

\begin{figure}
  \begin{center}
    \leavevmode
    \epsfysize=10cm
    \epsffile{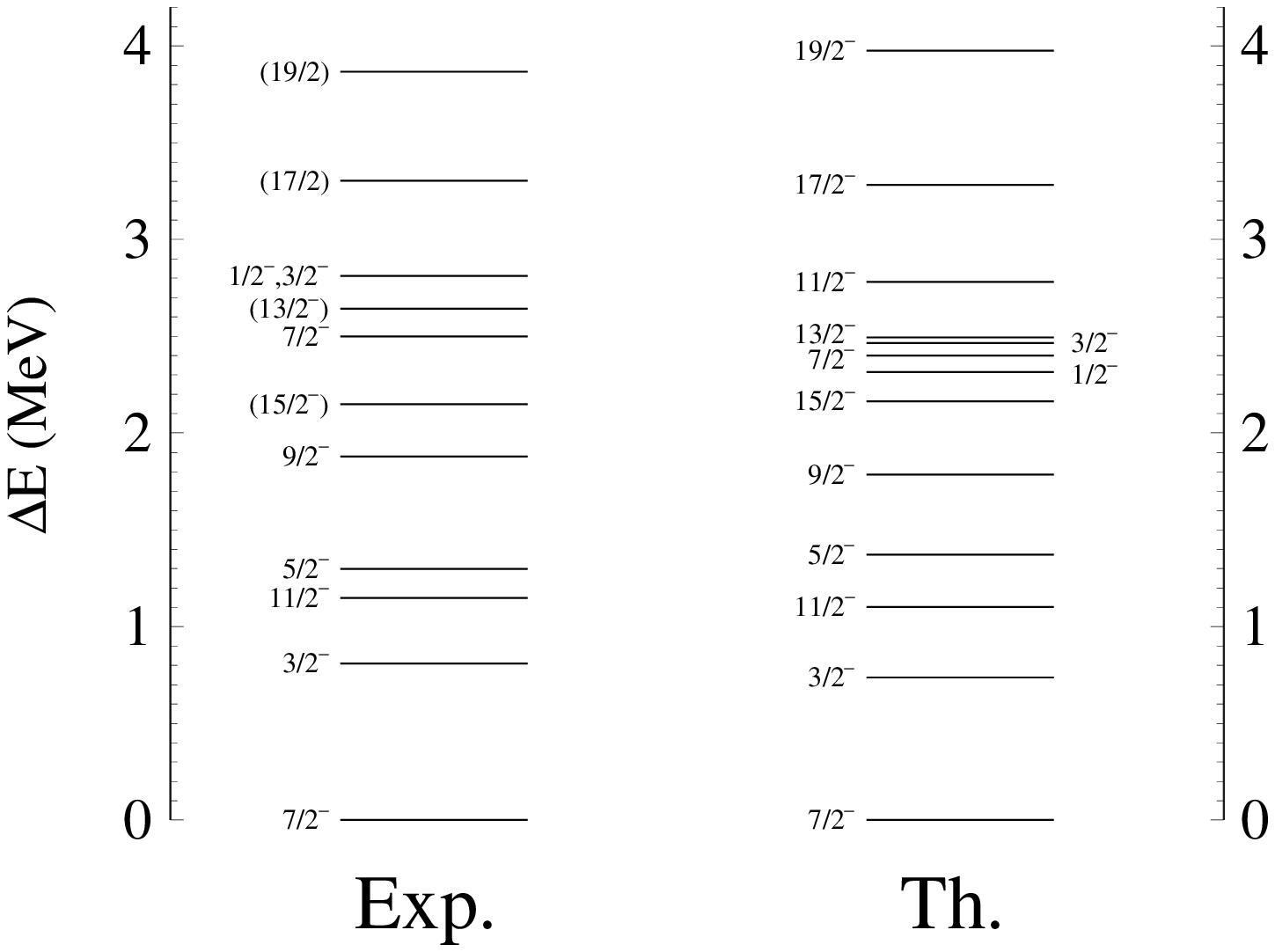}
    \caption{Theoretical and experimental energy levels of $^{47}$Sc.}
    \label{fig:e_sc47}
  \end{center}
\end{figure}

\begin{figure}
  \begin{center}
    \leavevmode
    \epsfysize=10cm
    \epsffile{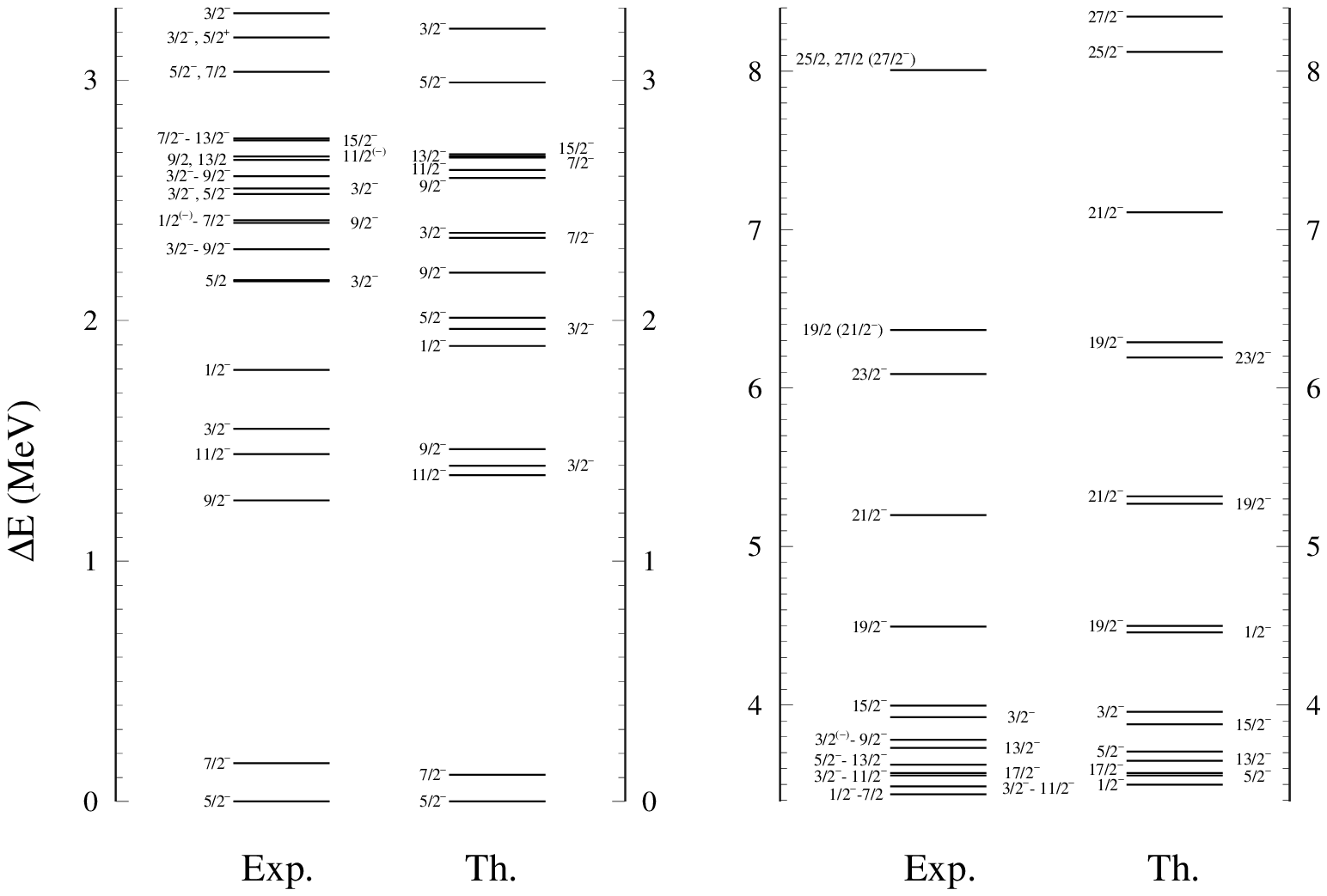}
    \caption{Theoretical and experimental energy levels of $^{47}$Ti.}
    \label{fig:e_ti47}
  \end{center}
\end{figure}

\begin{figure}
  \begin{center}
    \leavevmode
    \epsfysize=10cm
    \epsffile{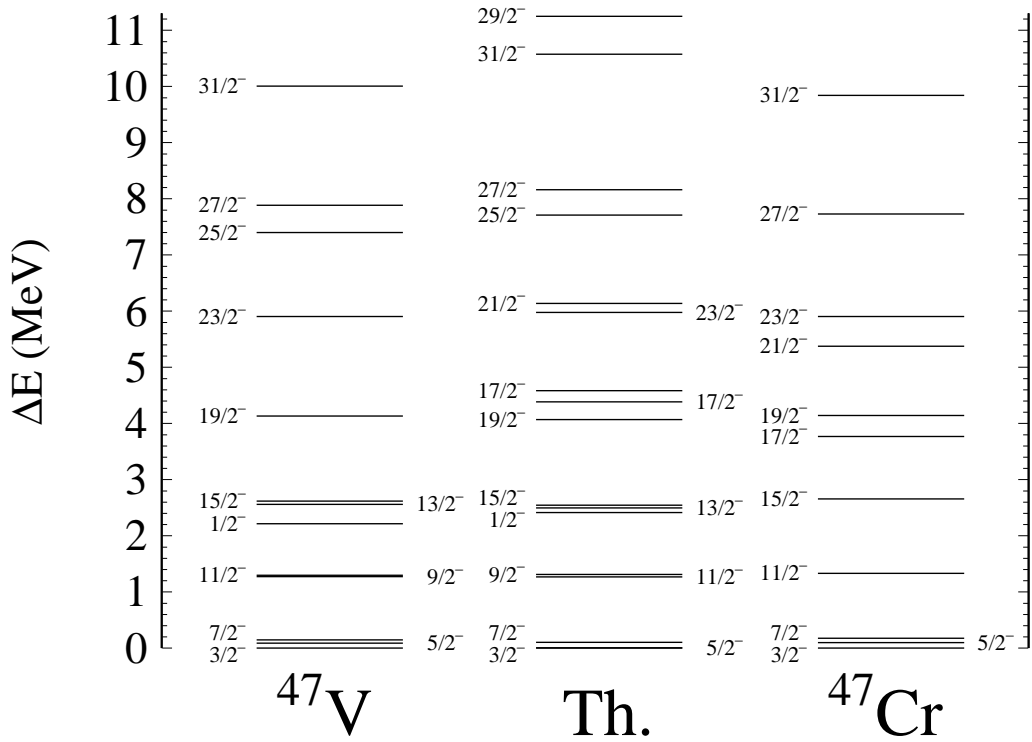}
    \caption{Energy levels of $^{47}$V and $^{47}$Cr.}
    \label{fig:e_crv47}
  \end{center}
\end{figure}

\begin{figure}
  \begin{center}
    \leavevmode
    \epsfysize=10cm
    \epsffile{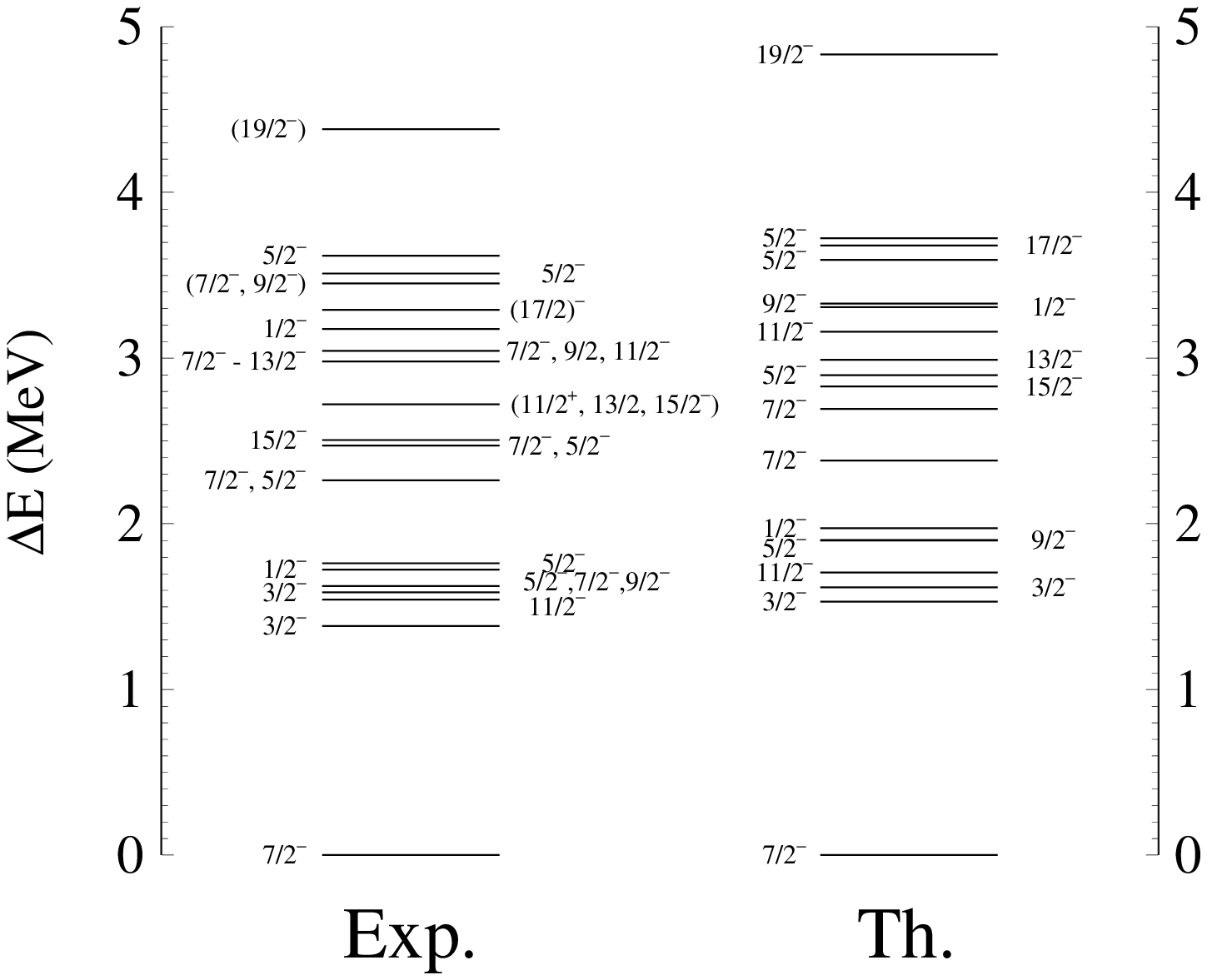}
    \caption{Theoretical and experimental energy levels of $^{49}$Ti.}
    \label{fig:e_ti49}
  \end{center}
\end{figure}

\begin{figure}
  \begin{center}
    \leavevmode
    \epsfysize=10cm
    \epsffile{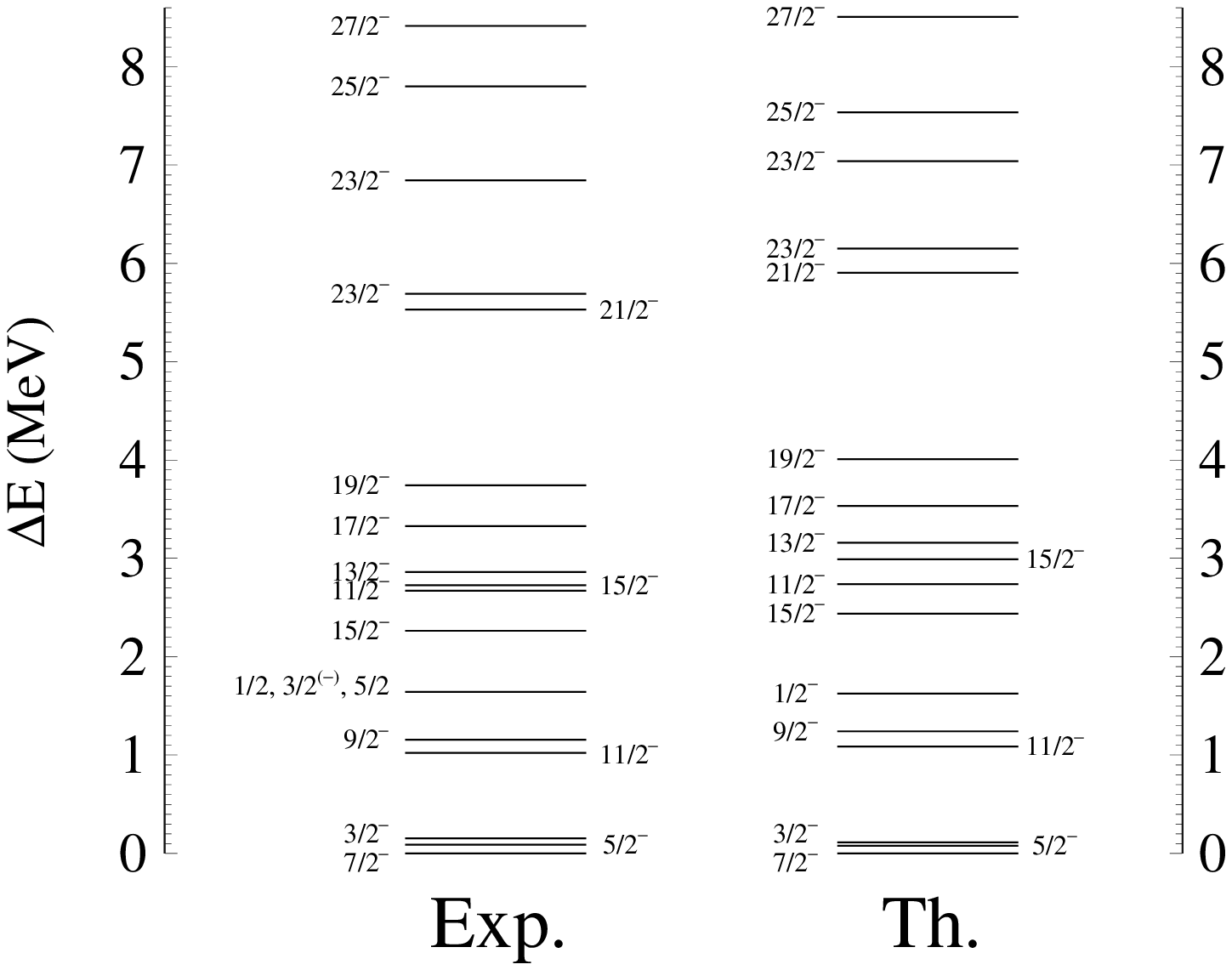}
    \caption{Theoretical and experimental energy levels of $^{49}$V.}
    \label{fig:e_v49}
  \end{center}
\end{figure}

\begin{figure}
  \begin{center}
    \leavevmode
    \epsfysize=6cm
    \epsffile{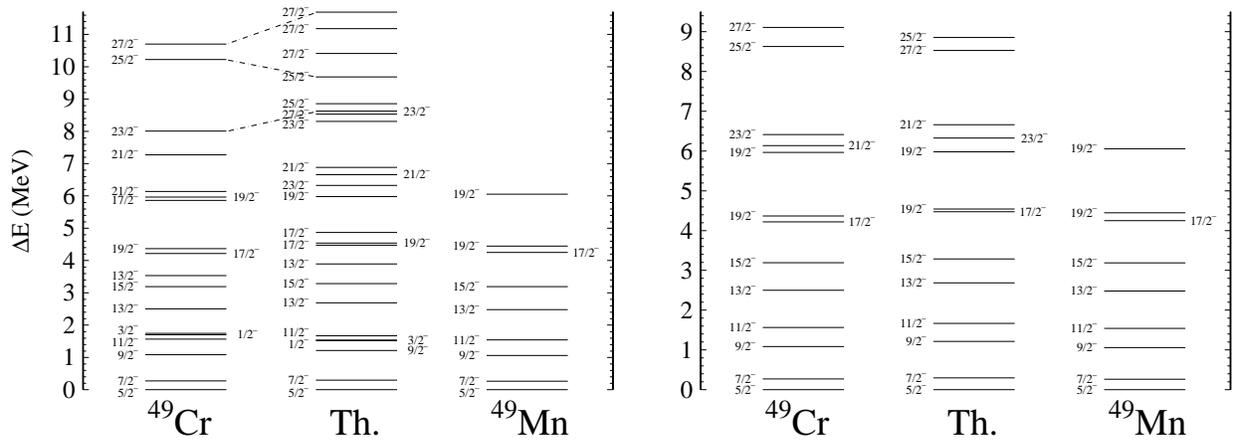}
    \caption{Energy levels of $^{49}$Cr and $^{49}$Mn. To the left
      comparison with experimental data as given in
      ref.~\protect\cite{cameron_49}. To the right comparison with our
      reinterpretation of these data as explained in the text.}
    \label{fig:e_crmn49}
  \end{center}
\end{figure}

\begin{figure}
  \begin{center}
    \leavevmode
    \epsfysize=10cm
    \epsffile{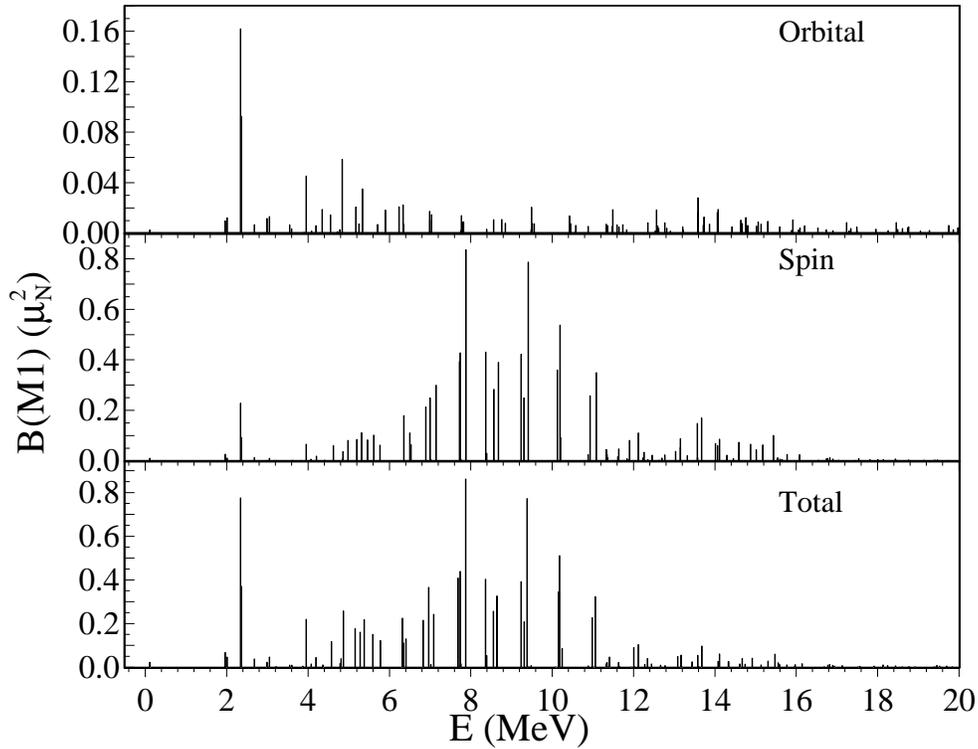}
    \caption{$M1$ strength functions of $^{47}$Ti, using bare
      gyromagnetic factors. 60 Lanczos iterations for each $J$}
    \label{fig:ti47m1}
  \end{center}
\end{figure}

\begin{figure}
  \begin{center}
    \leavevmode
    \epsfysize=8cm
    \epsffile{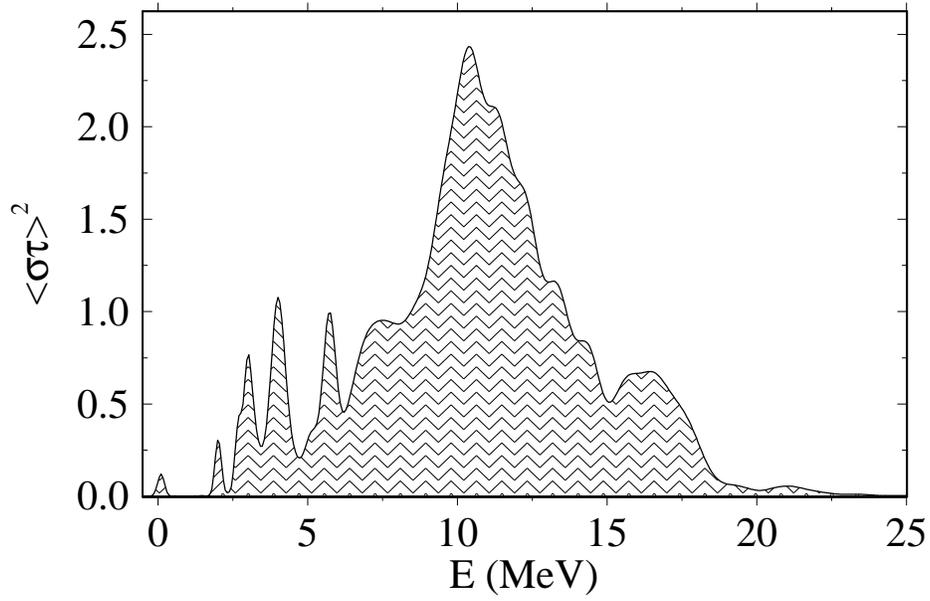}
    \caption{$^{47}$Sc$(\beta^-){}^{47}$Ti and
      $^{47}$Fe$(\beta^+){}^{47}$Mn strength function.}
    \label{fig:sc47beta}
  \end{center}
\end{figure}

\begin{figure}
  \begin{center}
    \leavevmode
    \epsfysize=12cm
    \epsffile{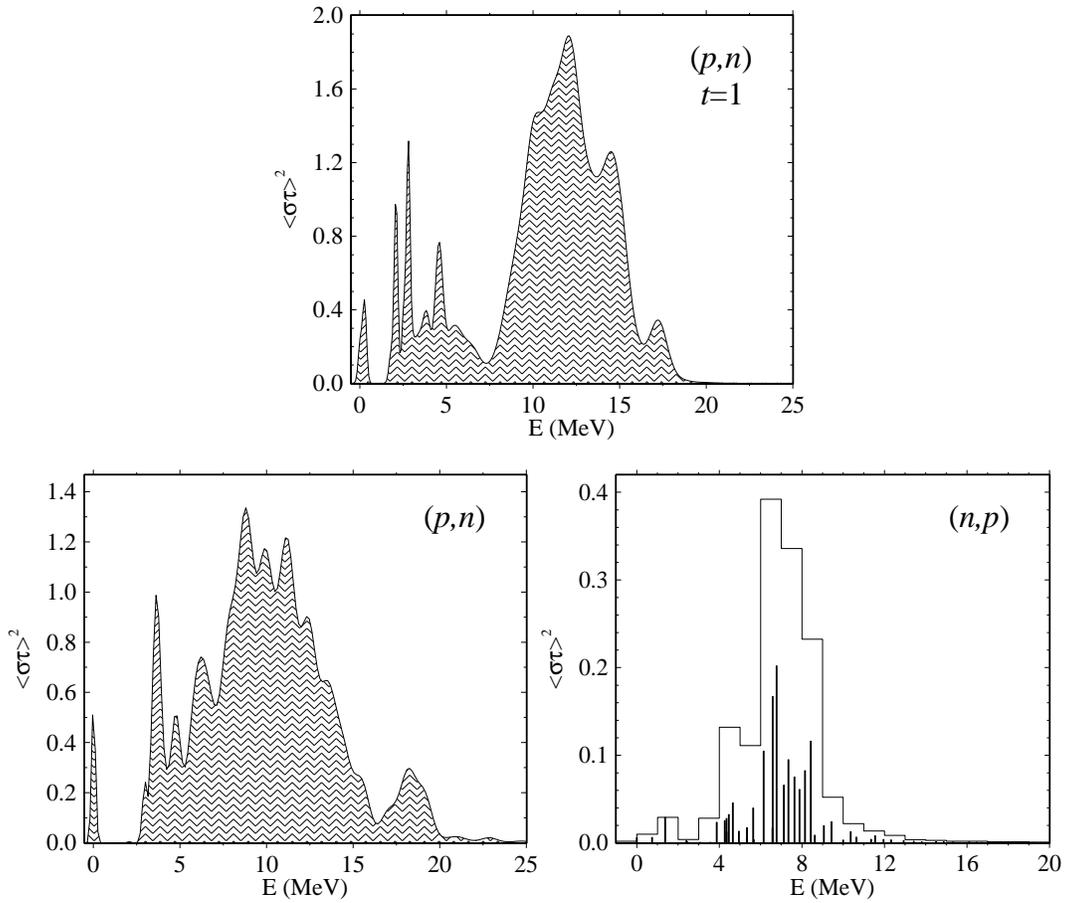}
    \caption{$^{47}$Ti$(p,n){}^{47}$V and $^{47}$Ti$(n,p){}^{47}$Sc
    strength functions.}
    \label{fig:ti47gt}
  \end{center}
\end{figure}

\begin{figure}
  \begin{center}
    \leavevmode
    \epsfysize=12cm
    \epsffile{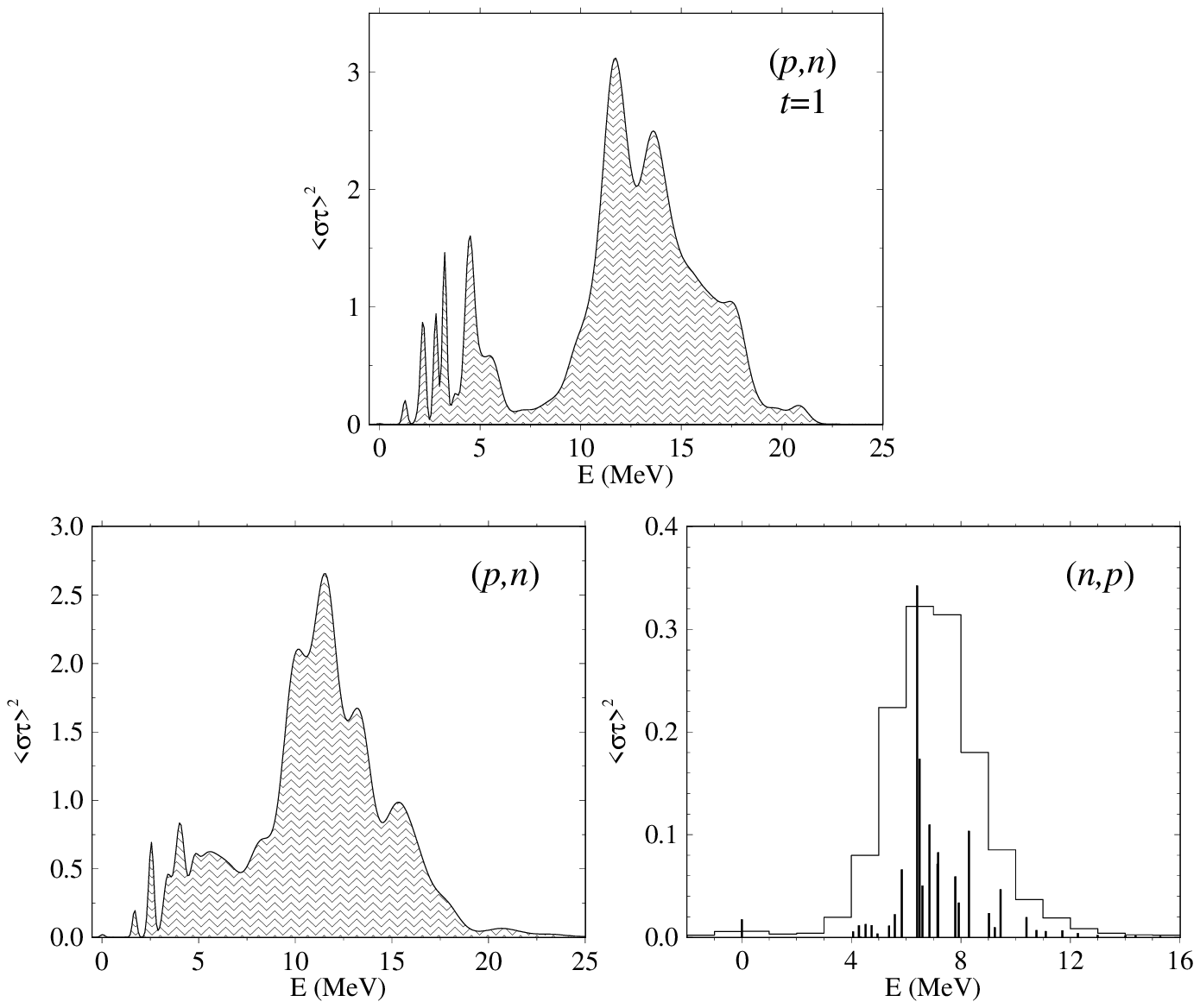}
    \caption{$^{49}$Ti$(p,n){}^{49}$V and $^{49}$Ti$(n,p){}^{49}$Sc
    strength functions.}
    \label{fig:ti49gt}
  \end{center}
\end{figure}

\begin{figure}
  \begin{center}
    \leavevmode
    \epsfysize=8cm
    \epsffile{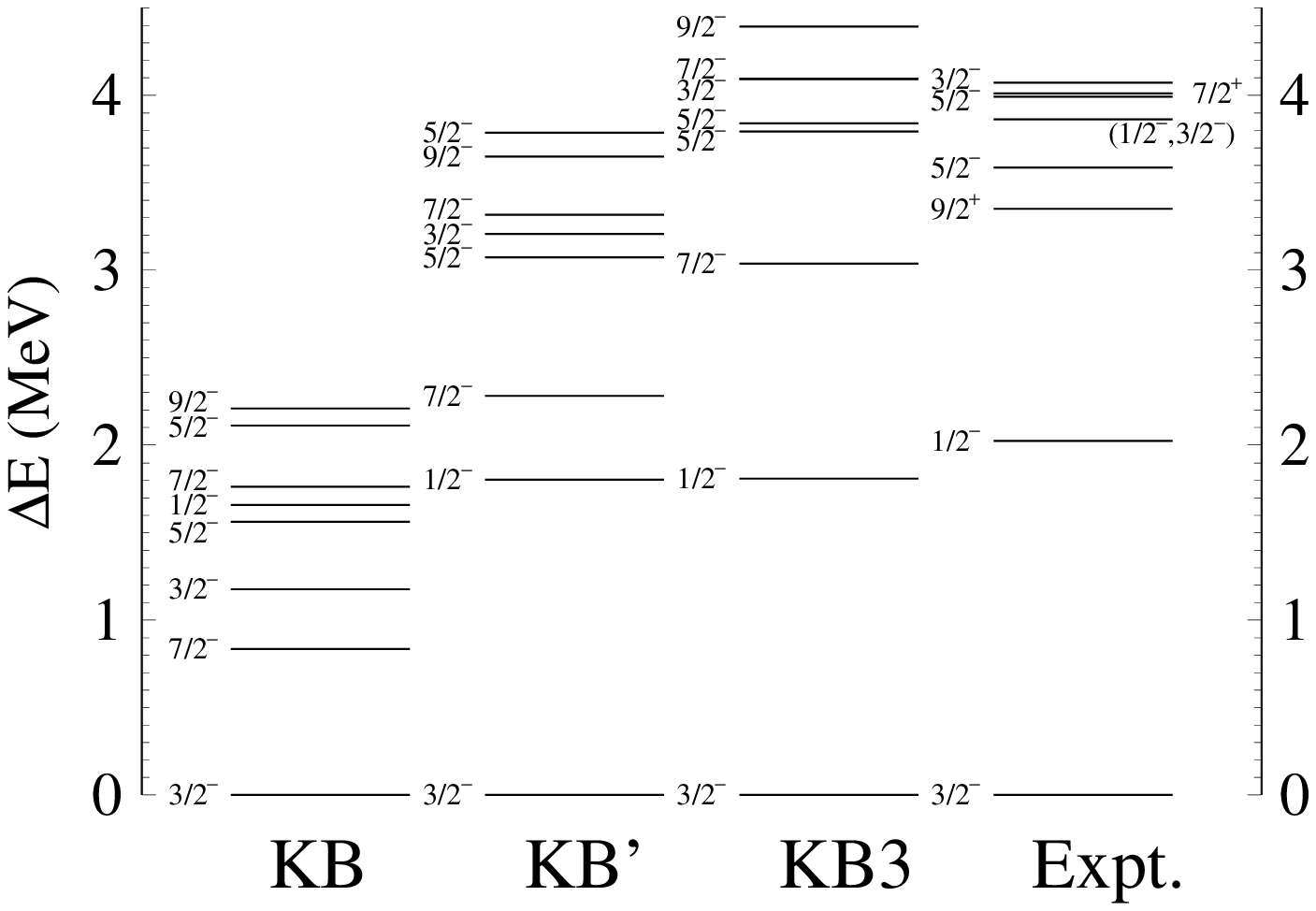}
    \caption{Theoretical levels for the KB, KB' and KB3 interactions
    compared with experiment in $^{49}$Ca.}
    \label{fig:e_ca49}
  \end{center}
\end{figure}

\begin{figure}
  \begin{center}
    \leavevmode
    \epsfysize=10cm
    \epsffile{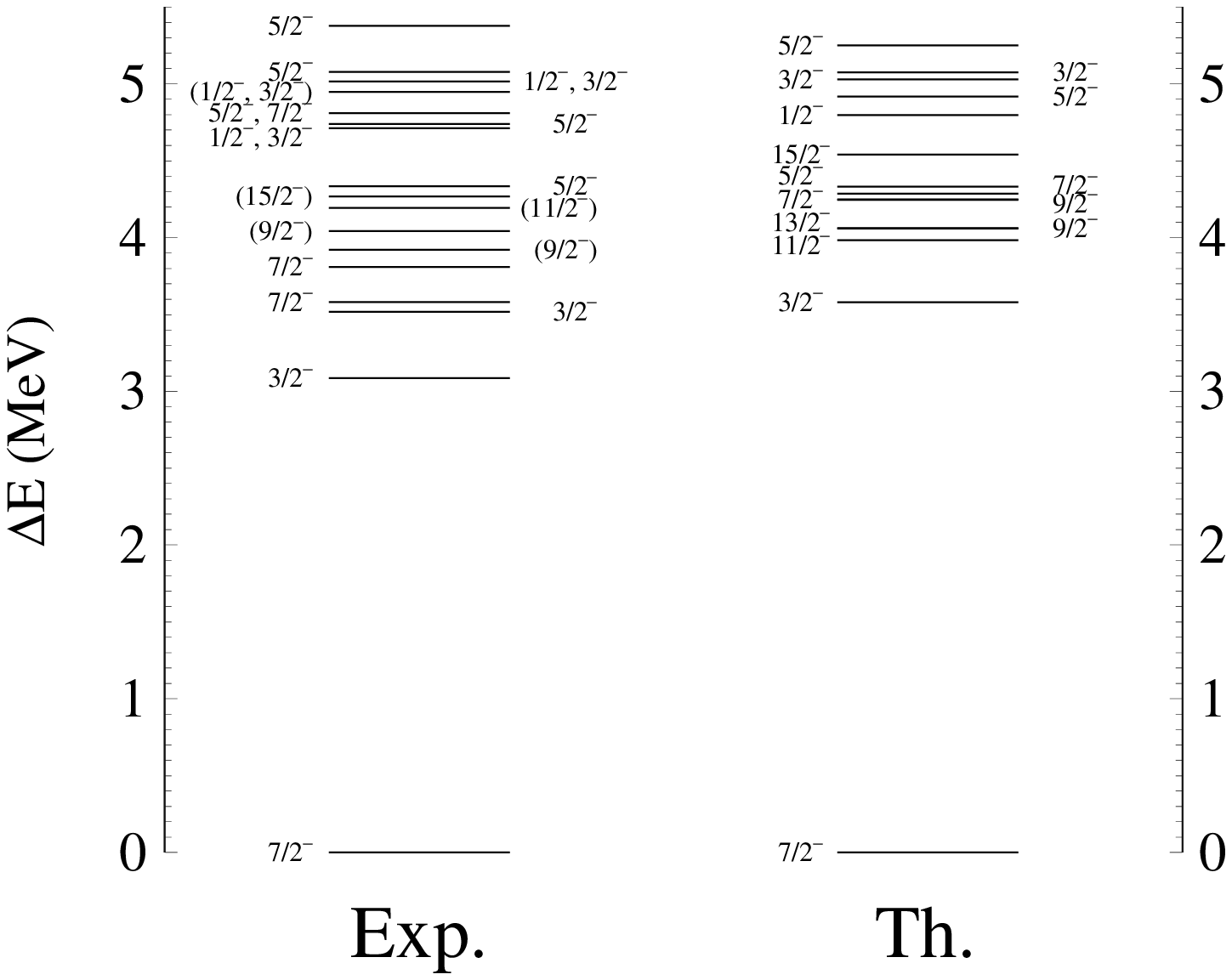}
    \caption{Theoretical and experimental energy levels of $^{49}$Sc.}
    \label{fig:e_sc49}
  \end{center}
\end{figure}

\begin{figure}
  \begin{center}
    \leavevmode
    \epsfysize=10cm
    \epsffile{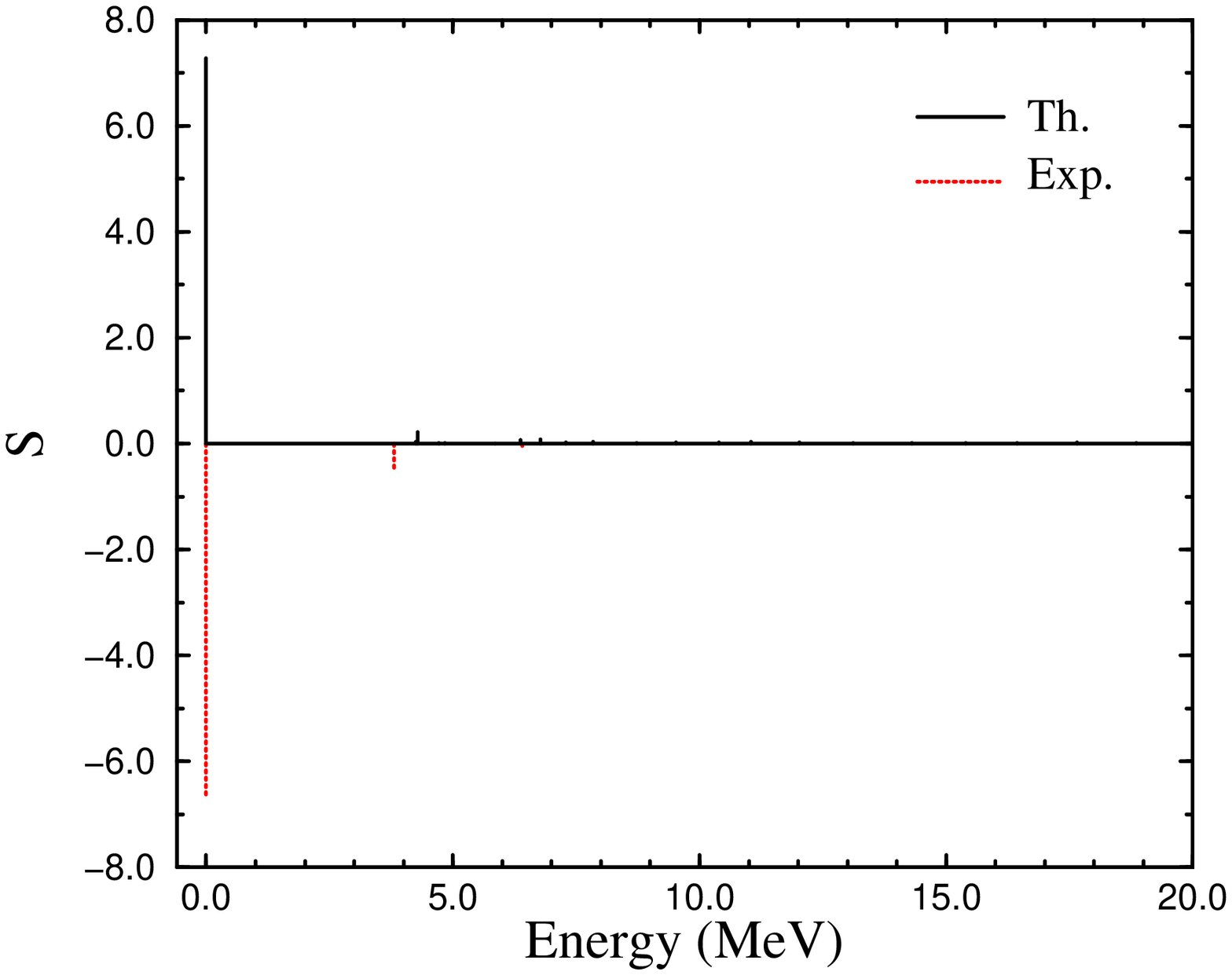}
    \caption{Spectroscopic factors, $(2j+1) S(j,t_z)$, corresponding
      to stripping of a particle in the orbit 1$f_{7/2}$.}
    \label{fig:spec_f7}
  \end{center}
\end{figure}

\begin{figure}
  \begin{center}
    \leavevmode
    \epsfysize=10cm
    \epsffile{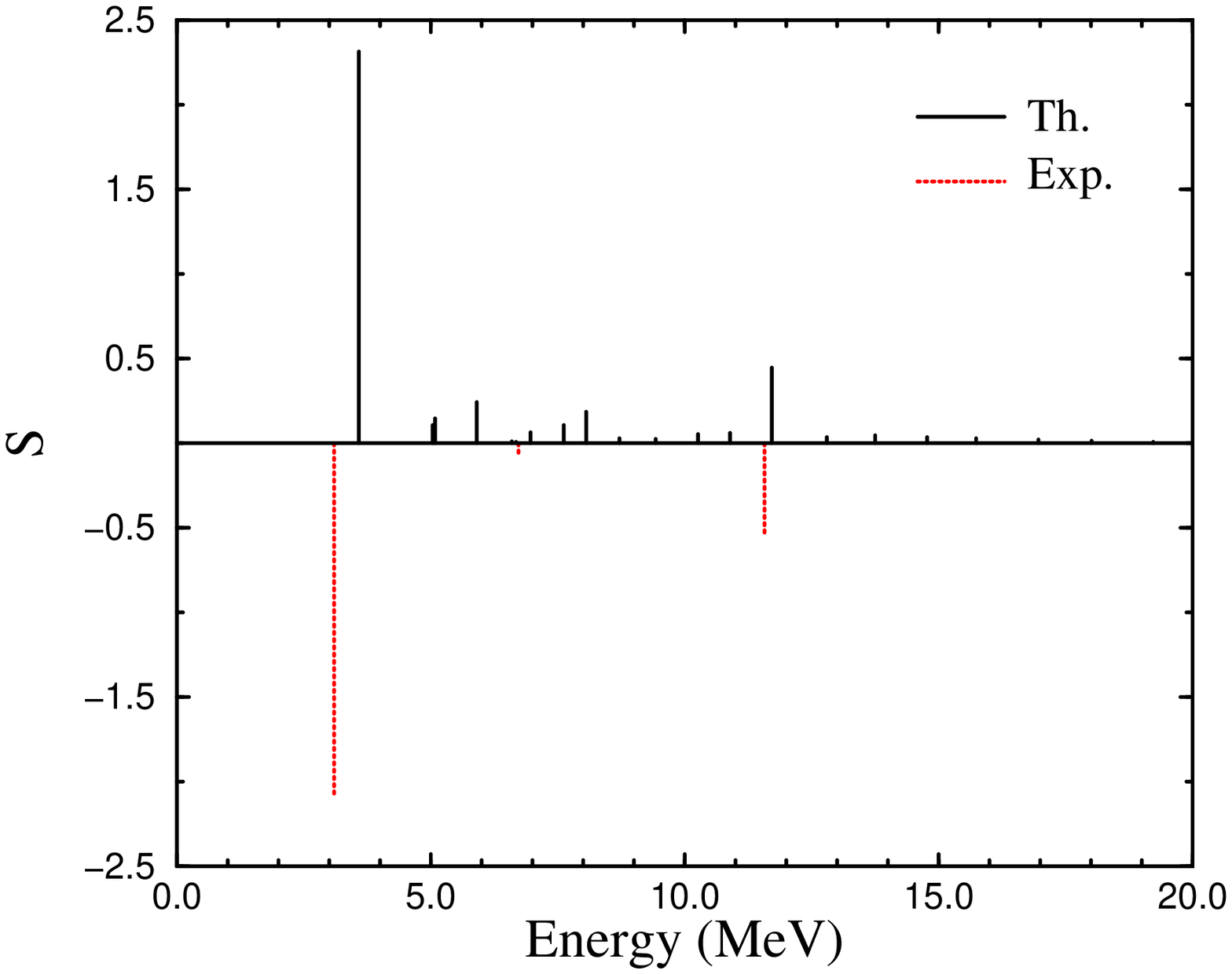}
    \caption{Spectroscopic factors, $(2j+1) S(j,t_z)$, corresponding
      to stripping of a particle in the orbit 0$p_{3/2}$.}
    \label{fig:spec_p3}
  \end{center}
\end{figure}

\begin{figure}
  \begin{center}
    \leavevmode
    \epsfysize=10cm
    \epsffile{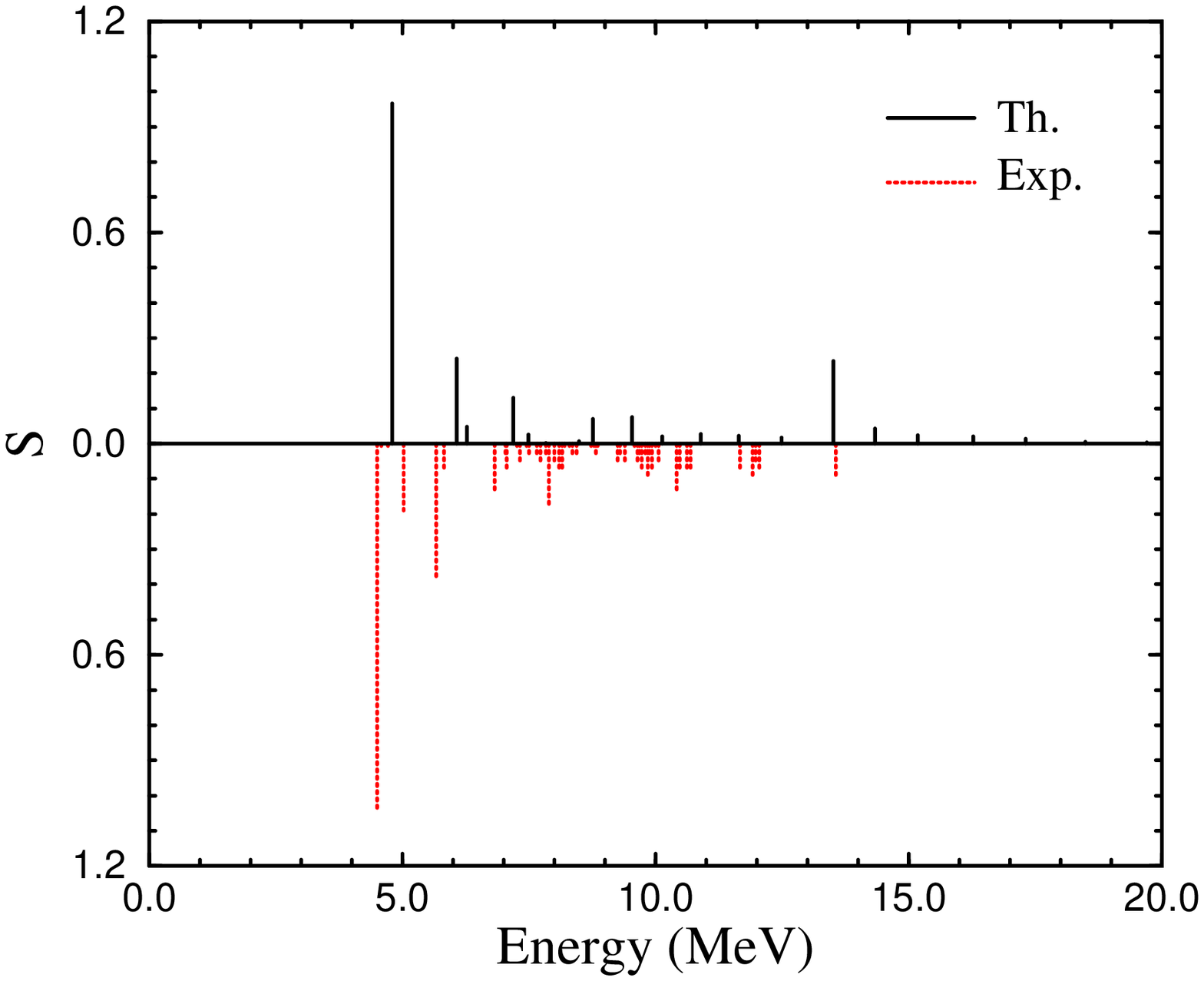}
    \caption{Spectroscopic factors, $(2j+1) S(j,t_z)$, corresponding
      to stripping of a particle in the orbit 0$p_{1/2}$.}
    \label{fig:spec_p1}
  \end{center}
\end{figure}

\begin{figure}
  \begin{center}
    \leavevmode
    \epsfysize=10cm
    \epsffile{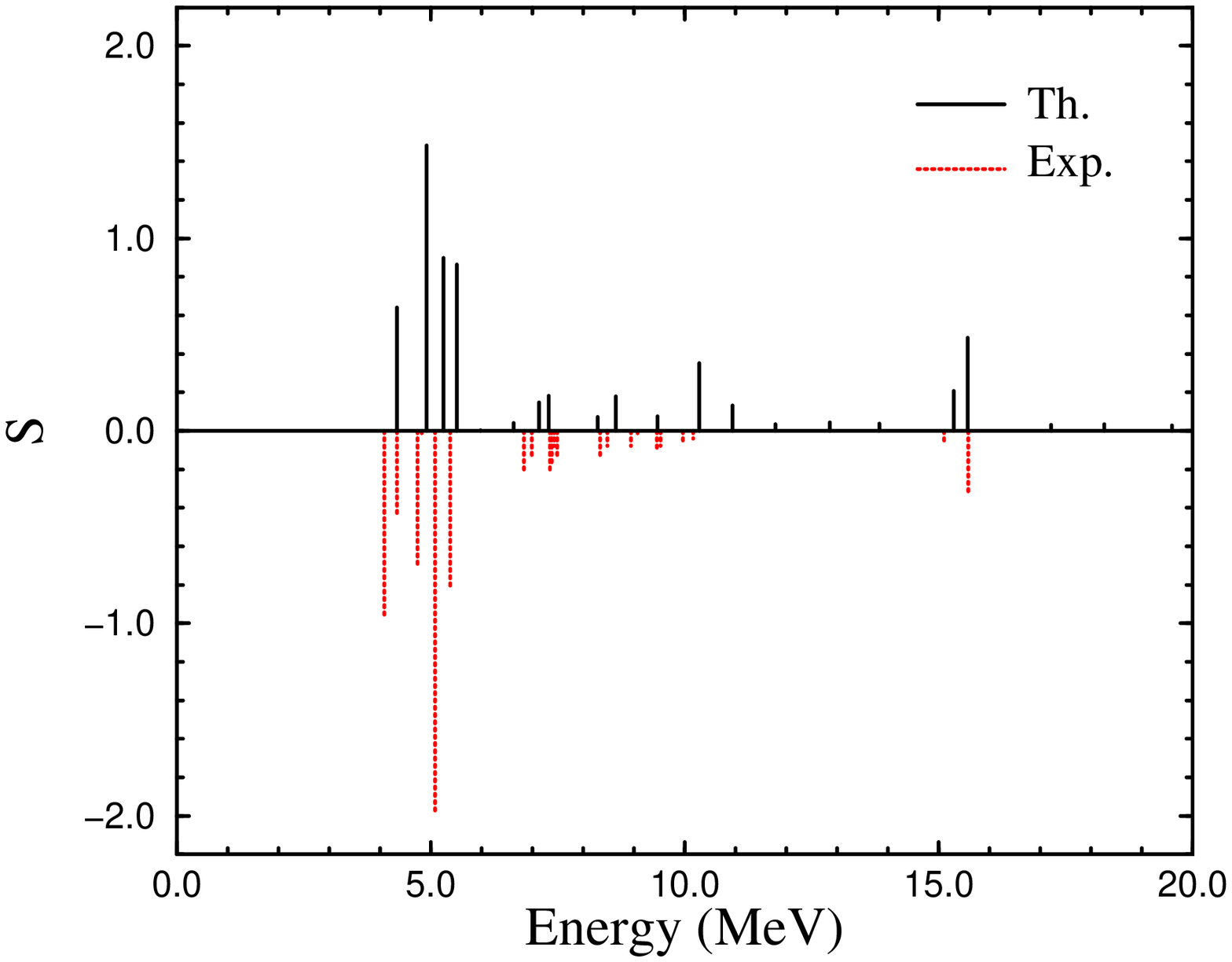}
    \caption{Spectroscopic factors, $(2j+1) S(j,t_z)$, corresponding
      to stripping of a particle in the orbit 1$f_{5/2}$.}
    \label{fig:spec_f5}
  \end{center}
\end{figure}

\begin{figure}
  \begin{center}
    \leavevmode
    \epsfysize=10cm
    \epsffile{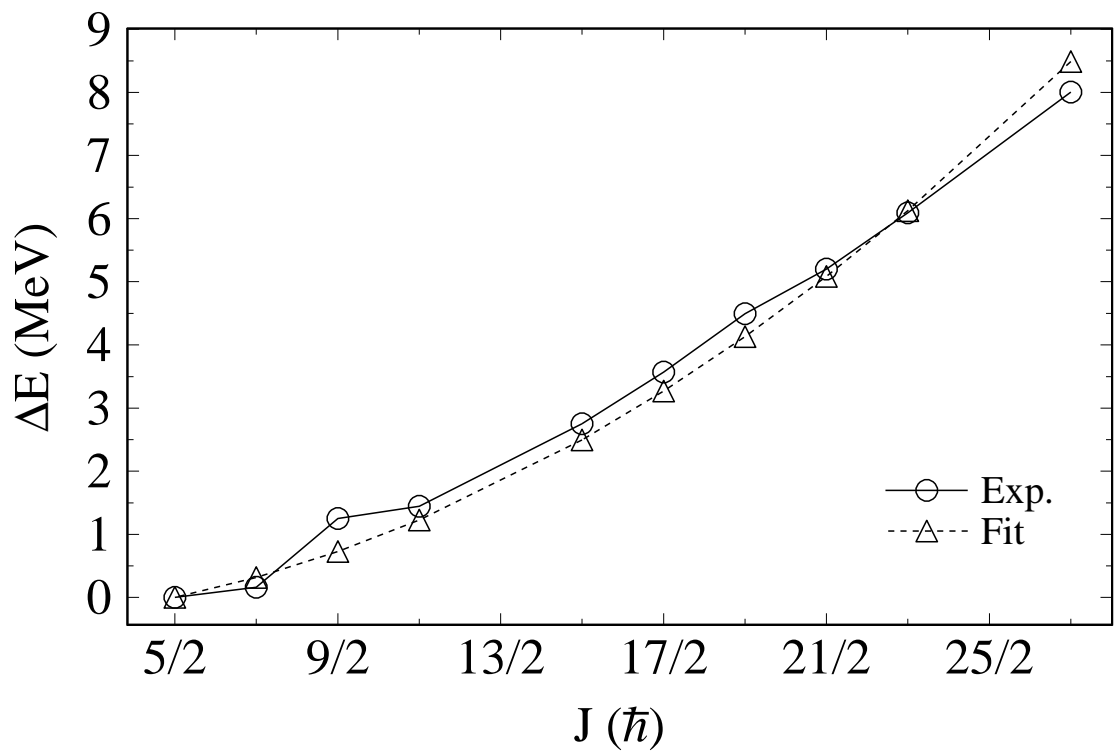}
    \caption{Experimental excitation energies of $^{47}$Ti compared
      with a rotational spectrum.}
    \label{fig:ti47rot}
  \end{center}
\end{figure}

\begin{figure}
  \begin{center}
    \leavevmode
    \epsffile{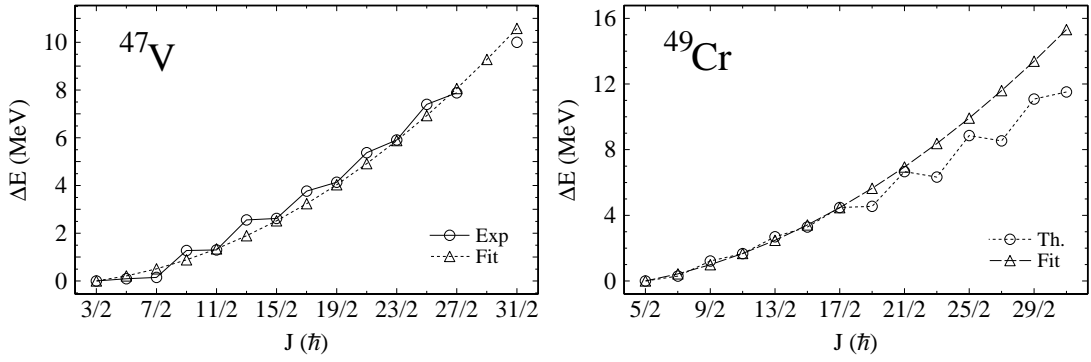}
    \caption{Comparison between the spectrum of $^{47}$V and $^{49}$Cr
      and the predictions of the rotational model. For $^{47}$V we use
      the experimental energies measured in~\protect\cite{cameron_47}
      with the energies of the 17/2 and 21/2 taken from the mirror
      nucleus $^{47}$Cr. For $^{49}$Cr we use the theoretical energies
      because of the discrepancies noted in
      section~\protect\ref{sec:el_49}}
    \label{fig:fit_47-49}
  \end{center}
\end{figure}

\begin{figure}
  \begin{center}
    \leavevmode
    \epsfysize=8cm
    \epsffile{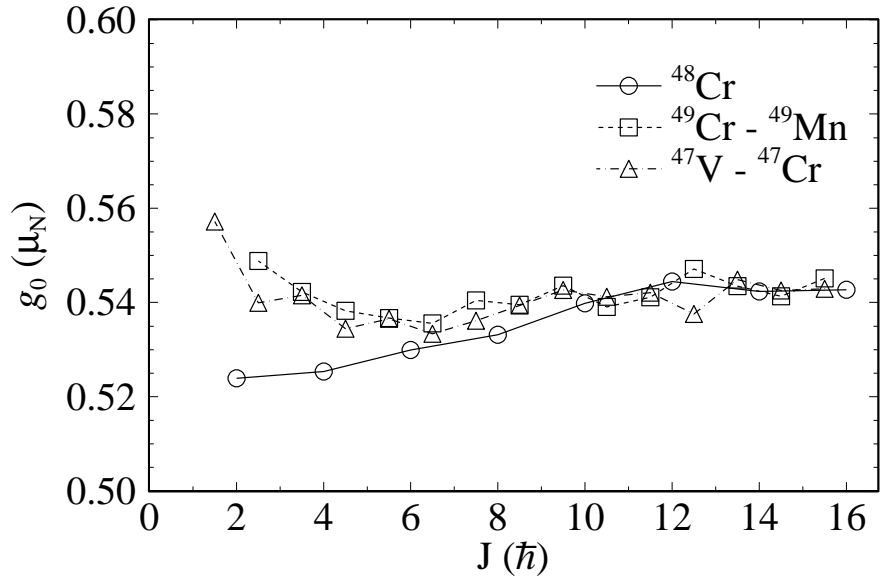}
    \caption{Gyromagnetics factors of $^{48}$Cr and its comparison
    with the isoscalar part of the gyromagnetic factors of the mirror
    pairs $^{47}$V-$^{47}$Cr and $^{49}$Cr-$^{49}$Mn.}
    \label{fig:g0}
  \end{center}
\end{figure}

\begin{figure}
  \begin{center}
    \leavevmode
    \epsffile{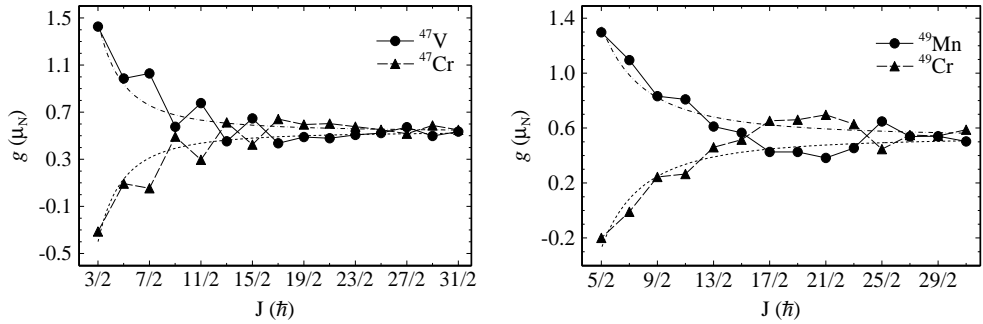}
    \caption{Gyromagnetics factors of the nuclei $^{47}$V, $^{47}$Cr,
    $^{49}$Cr and $^{49}$Mn as a function of the angular momentum. The
    curves represent the predictions of the particle plus rotor
    model.}
    \label{fig:47-49gr}
  \end{center}
\end{figure}

\begin{figure}
  \begin{center}
    \leavevmode
    \epsfysize=10cm
    \epsffile{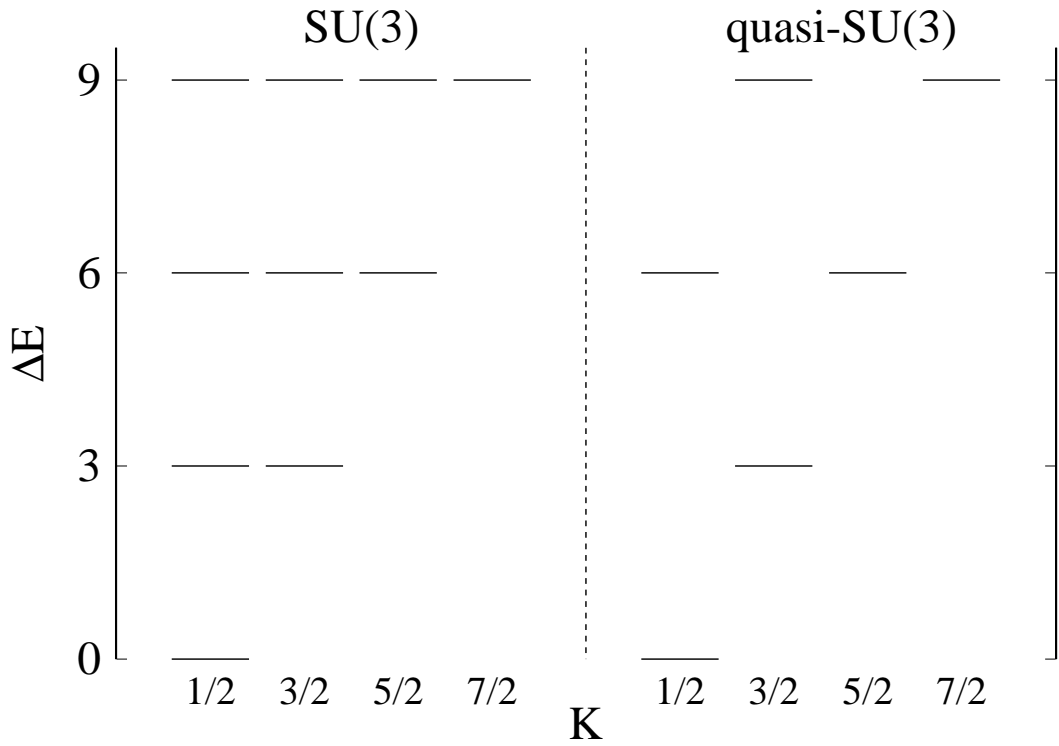}
    \caption{Nilsson orbits for SU(3) and quasi-SU(3). The band heads
    are at $-2p=-6$ for SU(3) and $-2p+1/2=-11/2$ for quasi-SU(3).}
    \label{fig:q-su3}
  \end{center}
\end{figure}

\begin{figure}
  \begin{center}
    \leavevmode
    \epsfxsize=16.5cm
    \epsffile{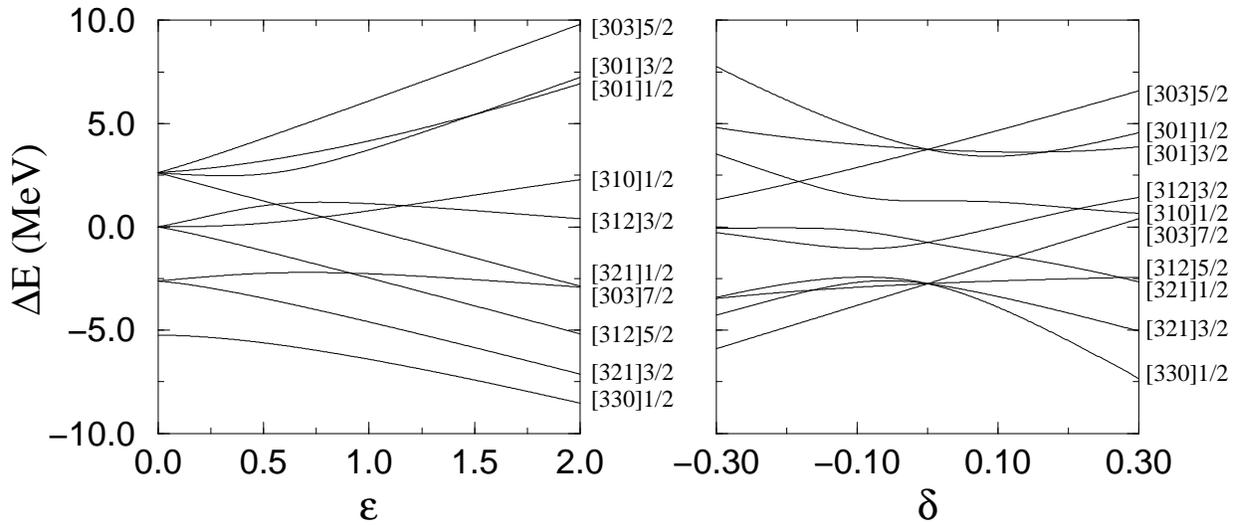}
    \caption{Nilsson diagrams in the $pf$ shell. Energy vs. single
      particle splitting $\varepsilon$ (left panel), energy vs.
      deformation $\delta$ (right panel).}
    \label{fig:nilsson}
  \end{center}
\end{figure}

\clearpage

\renewcommand{\arraystretch}{1.2}

\begin{table}
  \begin{center}
    \leavevmode
    \caption{$m$-scheme and maximal $JT$ dimensions in the full $pf$
      shell.}
    \label{tab:dime}
    \begin{tabular}{cccccc}
      & $^{47}$Ca & $^{47}$Sc & $^{47}$Ti & $^{47}$V & \\ \hline
      $m$-scheme & 7\,531 & 71\,351 & 262\,231 & 483\,887 & \\
      $2J\ 2T$ & 7 7 & 9 5 &  9 3 & 9 1 & \\
            & 1\,121 & 8\,858 & 25\,600 & 29\,121 & \\ \hline
      & $^{49}$Ca & $^{49}$Sc & $^{49}$Ti & $^{49}$V & $^{49}$Cr \\
      \hline
      $m$-scheme & 15\,666 & 219\,781 & 1\,227\,767 & 3\,580\,369 &
      6\,004\,205 \\
      $2J\ 2T$ & 7 9 & 9 7 & 9 5 & 9 3 & 9 1 \\
         & 2\,215 & 27\,091 & 127\,406 & 287\,309 & 289\,959
    \end{tabular}
  \end{center}
\end{table}

\begin{table}
  \begin{center}
    \leavevmode
    \caption{Dipole magnetic moments and quadrupole electric moments
      of the $A=47$ isobars.}
    \label{tab:muq47}
    \begin{tabular}{cccccc}
      Nucleus & State & \multicolumn{2}{c}{$\mu$ ($\mu_N$)} &
      \multicolumn{2}{c}{$Q$ ($e$ fm$^2$)} \\
      \cline{3-4}\cline{5-6}
      & & Expt. & Theor. & Expt. & Theor. \\
      \hline
      $^{47}$Ca & 7/2$^-$ (g.s.) & $-$1.380(24) & $-$1.41 & 2.1(4) & 6.7
      \\
      $^{47}$Sc & 7/2$^-$ (g.s.) & 5.34(2) & 5.12 & $-$22(3) & $-$21
      \\
      $^{47}$Ti & 5/2$^-$ (g.s.) & $-$0.78848(1) & $-$0.97 & 30.3(24)
      & 22.7 \\
      & 5/2$^-$ & $-$1.9(6) & $-$1.16 &  & 8.16\\
      $^{47}$V & 3/2$^-$ (g.s.) & & 2.14 & & 19.9 \\
      $^{47}$Cr & 3/2$^-$ (g.s.) & & $-$0.47 & & 20.6
    \end{tabular}
  \end{center}
\end{table}

\begin{table}
  \begin{center}
    \leavevmode
    \caption{$E2$ and $M1$ transitions of $^{47}$Sc.}
    \label{tab:t_sc47}
    \begin{tabular}{cccccc}
      $J^\pi_n$(i) & $J^\pi_m$(f) & \multicolumn{2}{c}{$B(E2)$ ($e^2$
        fm$^4$)} & \multicolumn{2}{c}{$B(M1)$ ($\mu_N^2$)}\\
      \cline{3-4}\cline{5-6}
      & & Expt. & Theor. & Expt. & Theor. \\ \hline
      $\frac{3}{2}^-_1$ & $\frac{7}{2}^-_1$ & 111(30) & 60 & & \\
      $\frac{11}{2}^-_1$ & $\frac{7}{2}^-_1$ & 91(30) & 24 & & \\
      $\frac{5}{2}^-_1$ & $\frac{3}{2}^-_1$ & 1108(605) & 38 &
      0.41(14) & 1.82 \\
      $\frac{5}{2}^-_1$ & $\frac{7}{2}^-_1$ & 0.9$^{+1.5}_{-0.9}$ &
      0.48 & 0.27(9) & 0.66 \\
      $\frac{9}{2}^-_1$ & $\frac{11}{2}^-_1$ & 130$^{+220}_{-130}$ &
      6.3 & 0.25(14) & 1.72 \\
      $\frac{9}{2}^-_1$ & $\frac{7}{2}^-_1$ & 3$^{+4}_{-3}$ & 9.2 &
      0.034(18) & 0.25 \\
      $\frac{7}{2}^-_2$ & $\frac{5}{2}^-_1$ & & & $>0.077$ & 1.17 \\
      $\frac{5}{2}^-_2$ & $\frac{3}{2}^-_1$ & & & $>0.016$ & 0.021
    \end{tabular}
  \end{center}
\end{table}

\begin{table}
  \begin{center}
    \leavevmode
    \caption{$E2$ and $M1$ transitions of $^{47}$Ti.}
    \label{tab:t_ti47}
    \begin{tabular}{cccccc}
      $J^\pi_n$(i) & $J^\pi_m$(f) & \multicolumn{2}{c}{$B(E2)$ ($e^2$
        fm$^4$)} & \multicolumn{2}{c}{$B(M1)$ ($\mu_N^2$)}\\
      \cline{3-4}\cline{5-6}
      & & Expt. & Theor. & Expt. & Theor. \\ \hline
      $\frac{7}{2}^-_1$ & $\frac{5}{2}^-_1$ & 252(50) & 140 &
      0.0460(14) & 0.0175 \\
      $\frac{9}{2}^-_1$ & $\frac{7}{2}^-_1$ & 191(40) & 102 & 0.188(20)
      & 0.154 \\
      $\frac{9}{2}^-_1$ & $\frac{5}{2}^-_1$ & 70(30) & 55 & & \\
      $\frac{11}{2}^-_1$ & $\frac{9}{2}^-_1$ & 705(605) & 83 &
      0.70(13) & 0.242 \\
      $\frac{11}{2}^-_1$ & $\frac{7}{2}^-_1$ & 159(25) & 98 & & \\
      $\frac{3}{2}^-_1$ & $\frac{7}{2}^-_1$ & 39(11) & 46 & & \\
      $\frac{3}{2}^-_1$ & $\frac{5}{2}^-_1$ & 3.3(15) & 22 &
      0.00270(90) & 0.00011 \\
      $\frac{1}{2}^-_1$ & $\frac{5}{2}^-_1$ & $<17$ & 21 & & \\
      $\frac{3}{2}^-_2$ & $\frac{7}{2}^-_1$ & 36.3(70) & 8.3 & & \\
      $\frac{3}{2}^-_2$ & $\frac{5}{2}^-_1$ & $<272$ & 0.48 &
      0.0734(125) & 0.102 \\
      $\frac{15}{2}^-_1$ & $\frac{11}{2}^-_1$ & 135(26) & 111 & & \\
      $\frac{17}{2}^-_1$ & $\frac{15}{2}^-_1$ & 604$^{+70}_{-604}$ &
      41 & 1.02(32) & 0.793 \\
      $\frac{19}{2}^-_1$ & $\frac{17}{2}^-_1$ & $<50$ & 25 &
      0.45(13) & 1.03
    \end{tabular}
  \end{center}
\end{table}

\begin{table}
  \begin{center}
    \leavevmode
    \caption{$E2$ and $M1$ transitions of $^{47}$V.}
    \label{tab:t_v47}
    \begin{tabular}{cccccc}
      $J^\pi_n$(i) & $J^\pi_m$(f) & \multicolumn{2}{c}{$B(E2)$ ($e^2$
        fm$^4$)} & \multicolumn{2}{c}{$B(M1)$ ($\mu_N^2$)}\\
      \cline{3-4}\cline{5-6}
      & & Expt. & Theor. & Expt. & Theor. \\ \hline
      $\frac{5}{2}^-_1$ & $\frac{3}{2}^-_1$ & 2418(806) & 251 &
      0.082(5) & 0.125 \\
      $\frac{7}{2}^-_1$ & $\frac{5}{2}^-_1$ & & & 0.354(43) & 0.239 \\
      $\frac{7}{2}^-_1$ & $\frac{3}{2}^-_1$ & 91(71) & 106 & & \\
      $\frac{9}{2}^-_1$ & $\frac{7}{2}^-_1$ & $161<B(E2)< 1300$ & 76
      & $0.0023<B(M1)<0.12$ & 0.080\\
      $\frac{9}{2}^-_1$ & $\frac{5}{2}^-_1$ & 181(60) & 138 & & \\
      $\frac{11}{2}^-_1$ & $\frac{7}{2}^-_1$ & 201(101) & 186 & &
    \end{tabular}
  \end{center}
\end{table}

\begin{table}
  \begin{center}
    \leavevmode
    \caption{Dipole magnetic moments and quadrupole electric moments
      of the $A=49$ isobars.}
    \label{tab:muq49}
    \begin{tabular}{cccccc}
      Nucleus & State & \multicolumn{2}{c}{$\mu$ ($\mu_N$)} &
      \multicolumn{2}{c}{$Q$ ($e$ fm$^2$)} \\
      \cline{3-4}\cline{5-6}
      & & Expt. & Theor. & Expt. & Theor. \\
      \hline
      $^{49}$Ca & 3/2$^-$ (g.s.) & $-$1.38(6)& $-$1.46 & & $-$3.95 \\
      $^{49}$Sc & 7/2$^-$ (g.s.) & & 5.38 & & $-$19.3 \\
      $^{49}$Ti & 7/2$^-$ (g.s.) & $-$1.10417(1) & $-$1.12 & 24(1)
      & 22 \\
      $^{49}$V & 7/2$^-$ (g.s.) & $\pm$4.47(5) & 4.37 & & $-$11.1\\
               & 3/2$^-$ (0.153) & 2.37(12) & 2.25 & & 18.87 \\
      $^{49}$Cr & 5/2$^-$ (g.s.) & $\pm$0.476(3) & $-$0.50 & & 36.1 \\
      & 19/2$^-$ (4.365) & 7.4(12) & 6.28 & & $-3.43$ \\
      $^{49}$Mn & 5/2$^-$ (g.s.) & & $-$3.24 & & 36.4
    \end{tabular}
  \end{center}
\end{table}

%
%

\begin{table}
  \begin{center}
    \leavevmode
    \caption{$E2$ and $M1$ transitions in $^{49}$Ti. The
      identification of the states follows the theoretical results.}
    \label{tab:t_ti49}
    \begin{tabular}{cccccc}
      $J^\pi_n$(i) & $J^\pi_m$(f) & \multicolumn{2}{c}{$B(E2)$ ($e^2$
        fm$^4$)} & \multicolumn{2}{c}{$B(M1)$ ($\mu_N^2$)}\\
      \cline{3-4}\cline{5-6}
      & & Expt. & Theor. & Expt. & Theor. \\ \hline
      $\frac{3}{2}^-_1$ & $\frac{7}{2}^-_1$ & 33(4) & 26 & & \\
      $\frac{11}{2}^-_1$ & $\frac{7}{2}^-_1$ & 65(6) & 72 & & \\
      $\frac{3}{2}^-_2$ & $\frac{7}{2}^-_1$ & $7.42<B(E2)<10.3$ & 54 & & \\
      $\frac{9}{2}^-_1$ & $\frac{7}{2}^-_1$ & $<1.49\times10^3$ & 101 &
      $<0.29$ & 0.24 \\
      $\frac{1}{2}^-_1$ & $\frac{3}{2}^-_1$ & $0.17<B(E2)<3.37\times10^3$
      & 125 & $1.39\times10^{-4}<B(M1)<2.72$ & 0.52 \\
      $\frac{1}{2}^-_1$ & $\frac{3}{2}^-_2$ & $32<B(E2)<7.0\times10^5$ &
      68 & $<0.93$ & 0.17 \\
      $\frac{5}{2}^-_1$ & $\frac{7}{2}^-_1$ & $11<B(E2)<202$ & 56 &
      $0.20<B(M1)<0.32$ & 0.31 \\
      $\frac{7}{2}^-_2$ & $\frac{5}{2}^-_1$ & & & 0.88(27) & 0.79 \\
      $\frac{7}{2}^-_2$ & $\frac{9}{2}^-_1$ & & & 0.98(45) & 1.07 \\
      $\frac{7}{2}^-_2$ & $\frac{7}{2}^-_1$ & $<73$ & 18 & $<0.27$ &
      0.0096 \\
      $\frac{7}{2}^-_3$ & $\frac{5}{2}^-_1$ & & & 0.38(18) & 0.71 \\
      $\frac{7}{2}^-_3$ & $\frac{9}{2}^-_1$ & & & 0.48(23) & 0.97 \\
      $\frac{7}{2}^-_3$ & $\frac{7}{2}^-_1$ & $<50$ & 11 & $<0.021$ &
      0.012 \\
      $\frac{15}{2}^-_1$ & $\frac{11}{2}^-_1$ & $0.099<B(E2)<195$ &
      46 & & \\
      $\frac{17}{2}^-_1$ & $\frac{15}{2}^-_1$ & $>270$ & 49 & $>1.15$
      & 1.74 \\
    \end{tabular}
  \end{center}
\end{table}

\begin{table}
  \begin{center}
    \leavevmode
    \caption{$E2$ and $M1$ transitions in $^{49}$V.}
    \label{tab:t_v49}
    \begin{tabular}{cccccc}
      $J^\pi_n$(i) & $J^\pi_m$(f) & \multicolumn{2}{c}{$B(E2)$ ($e^2$
        fm$^4$)} & \multicolumn{2}{c}{$B(M1)$ ($\mu_N^2$)}\\
      \cline{3-4}\cline{5-6}
      & & Expt. & Theor. & Expt. & Theor. \\ \hline
      $\frac{5}{2}^-_1$ & $\frac{7}{2}^-_1$ & & & 0.224(14) &
      0.12\\
      $\frac{3}{2}^-_1$ & $\frac{5}{2}^-_1$ & & & 0.00267(9) &
      0.0028 \\
      $\frac{3}{2}^-_1$ & $\frac{7}{2}^-_1$ & 204(6) & 196 & & \\
      $\frac{11}{2}^-_1$ & $\frac{7}{2}^-_1$ & 149(27) & 157 & & \\
      $\frac{9}{2}^-_1$ & $\frac{11}{2}^-_1$ & & & 0.41(18) & 0.50\\
      $\frac{9}{2}^-_1$ & $\frac{5}{2}^-_1$ & 84(24) & 88 & & \\
      $\frac{9}{2}^-_1$ & $\frac{7}{2}^-_1$ & 63(19) & 41 &
      0.0120(34) & 0.0066 \\
      $\frac{15}{2}^-_1$ & $\frac{11}{2}^-_1$ & 298$^{+85}_{-180}$ &
      140 & & \\
      $\frac{11}{2}^-_2$ & $\frac{11}{2}^-_1$ & $>460$ & 0.46 &
      $>0.66$ & 0.72 \\
      $\frac{15}{2}^-_2$ & $\frac{15}{2}^-_1$ & $<1.28(53)\times10^3$
      & 99 & 2.0(9) & 1.40 \\
      $\frac{15}{2}^-_2$ & $\frac{11}{2}^-_1$ & 190(90) & 34 & & \\
      $\frac{13}{2}^-_1$ & $\frac{15}{2}^-_1$ & & & 0.39(30) & 1.0 \\
      $\frac{13}{2}^-_1$ & $\frac{9}{2}^-_1$ & 290(200) & 110 & & \\ 
    \end{tabular}
  \end{center}
\end{table}

\begin{table}
  \begin{center}
    \leavevmode
    \caption{$E2$ and $M1$ transitions in $^{49}$Cr.}
    \label{tab:t_cr49}
    \begin{tabular}{cccccc}
      $J^\pi_n$(i) & $J^\pi_m$(f) & \multicolumn{2}{c}{$B(E2)$ ($e^2$
        fm$^4$)} & \multicolumn{2}{c}{$B(M1)$ ($\mu_N^2$)}\\
      \cline{3-4}\cline{5-6}
      & & Expt. & Theor. & Expt. & Theor. \\ \hline
      $\frac{7}{2}^-_1$ & $\frac{5}{2}^-_1$ & 383(117) & 332 & 0.15(4)
      & 0.14 \\
      $\frac{9}{2}^-_1$ & $\frac{7}{2}^-_1$ & 426(149) & 283 & 0.45(9)
      & 0.39 \\
      $\frac{9}{2}^-_1$ & $\frac{5}{2}^-_1$ & 149(43) & 97 & & \\
      $\frac{11}{2}^-_1$ & $\frac{9}{2}^-_1$ & 107(85) & 213 & 0.50(9)
      & 0.47 \\
      $\frac{11}{2}^-_1$ & $\frac{7}{2}^-_1$ & 213(43) & 166 & & \\
      $\frac{1}{2}^-_1$ & $\frac{5}{2}^-_1$ & $<10$ & 7.0 & & \\
      $\frac{3}{2}^-_1$ & $\frac{7}{2}^-_1$ & 21(6) & 4.1 & & \\
      $\frac{3}{2}^-_1$ & $\frac{5}{2}^-_1$ & 0.26(12) & 4.8 &
      0.0048(14) & 0.00003 \\
      $\frac{13}{2}^-_1$ & $\frac{11}{2}^-_1$ & $4^{+106}_{-4}$ & 153
      & 0.27(7) & 0.62 \\
      $\frac{13}{2}^-_1$ & $\frac{9}{2}^-_1$ & 64(30) & 192 & & \\
      $\frac{15}{2}^-_1$ & $\frac{13}{2}^-_1$ & $<256$ & 92 &
      0.20(7) & 0.79 \\
      $\frac{15}{2}^-_1$ & $\frac{11}{2}^-_1$ & 92(27) & 185 & & \\
      $\frac{19}{2}^-_1$ & $\frac{17}{2}^-_1$ & & & 0.356(32) &
      0.520\\
      $\frac{19}{2}^-_1$ & $\frac{15}{2}^-_1$ & 127(10) & 158 & & \\
    \end{tabular}
  \end{center}
\end{table}

\begin{table}
  \begin{center}
    \leavevmode
    \caption{$S_+$ strength for  $t=1$ and full calculations.
     $\overline q$ is the average of the ratios between the two values
     of the strength for a given $Z$}
    \label{tab:smas}
    \begin{tabular}{cccccc}
      Nucleus & Space & $A=47$ & $A=48$ & $A=49$ & $\overline q$ \\
      \hline
      Sc & $t=1$ & 1.72 & 1.71 & 1.71 & 1.96 \\
      & full & 0.89 & 0.85 & 0.88 & \\
      Ti & $t=1$ & 3.45 & 3.44 & 3.43 & 2.60 \\
      & full & 1.39 & 1.26 & 1.33 & \\
      V  & $t=1$ & 5.58 & 5.23 & 5.16 & 1.88 \\
      & full & 2.94 & 2.87 & 2.67 & \\
      Cr & $t=1$ &  & 7.66 & 7.24 & 1.80 \\
         & full &   & 4.13 & 4.13 &
    \end{tabular}
  \end{center}
\end{table}

\begin{table}
  \begin{center}
    \leavevmode
    \caption{Experimental and theoretical half-lives using the
      quenching factor 0.77.}
    \label{tab:half47}
    \begin{tabular}{ccccc}
      Nucleus & \multicolumn{2}{c}{Half-Life} &
      \multicolumn{2}{c}{Fermi (\%)}\\
      \cline{2-3} \cline{4-5}
      & Expt. & Theor. & Expt. & Theor. \\
      \hline
      $^{47}$Ca & 4.336(3) d & 4.20 d & & \\
      $^{47}$Sc & 3.3492(6) d & 3.79 d & & \\
      $^{47}$V & 36.6(3) m & 20.7 m & & \\
      $^{47}$Cr & 500(15) ms & 480 ms & 78.7 & 76.1 \\
      $^{47}$Mn &  & 65.2 ms & & 54.1 \\
      $^{47}$Fe & & 18.7 ms & & 26
    \end{tabular}
  \end{center}
\end{table}

\begin{table}
  \begin{center}
    \leavevmode
    \caption{Experimental and theoretical half-lifes using the
      quenching factor 0.77}
    \label{tab:half49}
    \begin{tabular}{ccccc}
      Nucleus & \multicolumn{2}{c}{Half-Life} &
      \multicolumn{2}{c}{Fermi (\%)}\\
      \cline{2-3} \cline{4-5}
      & Expt. & Theor. & Expt. & Theor. \\
      \hline
      $^{49}$Ca & 8.718(6) m & 3.17 m & & \\
      $^{49}$Sc & 57.2(2) m & 41.4 m & & \\
      $^{49}$V  & 330(15) d & 1088 d & & \\
      $^{49}$Cr & 42.3(1) m & 38.2 m & & \\
      $^{49}$Mn & 382(7) ms & 398 ms & 72 & 75 \\
      $^{49}$Fe & 75(10) & 55 ms & 61 & 42
    \end{tabular}
  \end{center}
\end{table}

\begin{table}
  \begin{center}
    \leavevmode
    \caption{Binding energies (in MeV) of the $A=47$ nuclei relative
      to $^{40}$Ca. The experimental values are from
      reference~\protect\cite{mass}. Asterisks are used for those
      obtained from systematics.}
    \label{tab:be_47}
    \begin{tabular}{lcccccdd}
      & Ca & Sc & Ti & V & Cr & Mn & Fe \\ \hline
      Expt. & 63.99 & 65.20 & 65.02 & 61.31 & 53.08 & 40.01$^*$ &
      23.58$^*$ \\ 
      Theor. & 64.06 & 65.11 & 64.81 & 61.07 & 52.89 & 40.28 & 24.22
    \end{tabular}
  \end{center}
\end{table}

\begin{table}
  \begin{center}
    \leavevmode
    \caption{Binding energies (in MeV) of the $A=49$ nuclei relative
      to $^{40}$Ca. The experimental values are from
      reference~\protect\cite{mass}. Asterisks are used for those
      obtained from systematics.}
    \label{tab:be_49}
    \begin{tabular}{lccccccdd}
      & Ca & Sc & Ti & V & Cr & Mn & Fe & Co\\ \hline
      Expt. & 79.09 & 83.57 & 84.79 & 83.40 & 79.99 & 71.50 &
      57.69$^*$ & 42.20$^*$ \\
      Theor. & 78.75 & 83.69 & 85.12 & 83.70 & 80.23 & 71.75 & 58.27 &
      42.73
    \end{tabular}
  \end{center}
\end{table}

\begin{table}
  \begin{center}
    \leavevmode
    \caption{Intrinsic quadrupole moments ($e$ fm$^2$) of
      $^{47}$Ti. The numbers in parenthesis have been calculated using
      the experimental data. $Q_0^{(s)}$ means computed from the
      spectroscopic moment. $Q_0^{(t)}$ from the $B(E2)$ value.}
    \label{tab:q0_ti47}
    \begin{tabular}{cccc}
      $J$ & $Q_0^{(s)}$ & \multicolumn{2}{c}{$Q_0^{(t)}$} \\
      \cline{3-4}
      &  & $\Delta J=1$ & $\Delta J =2$ \\ \hline
      $5/2^-$ & 63.4 (84.8) &  &  \\
      $7/2^-$ & 120.3 & 61.6 (84.2) & \\
      $9/2^-$ & 44.1  & 57.1 (79.6) & 72.7 (83.9) \\
      $11/2^-$ & 10.9 & 58.6 & 74.7 (97.1) \\
      $13/2^-$ & 60.7 & 57.3 & 67.5 \\
      $15/2^-$ & 29.8 & 50.2 & 66.0 (74.1) \\
      $17/2^-$ & 69.7 & 59.3 & 59.7 \\
      $19/2^-$ & 45.6 & 50.6 & 51.9 \\
      $21/2^-$ & 66.6 & 40.3 & 40.4 \\
      $23/2^-$ & 53.5 & 56.8 & 43.1 \\
      $25/2^-$ & 30.1 & 9.1 & 20.3 \\
      $27/2^-$ & 46.7 & 65.8 & 28.8 \\
    \end{tabular}
  \end{center}
\end{table}


\begin{table}
  \begin{center}
    \caption{Intrinsic quadrupole moments ($e$ fm$^2$) of the mirror
      pair $^{47}$V-$^{47}$Cr. The numbers in parenthesis have been
      calculated using the experimental data. $Q_0^{(s)}$ means
      computed from the spectroscopic moment. $Q_0^{(t)}$ from the
      $B(E2)$ value. When a state can decay to two, both numbers are
      given. }
    \label{tab:q0_47}
    \leavevmode
    \begin{tabular}{ccccccc}
      & \multicolumn{3}{c}{$^{47}$V} & \multicolumn{3}{c}{$^{47}$Cr}
      \\ 
      $J^\pi_n$ & $Q_0^{(s)}$ & \multicolumn{2}{c}{$Q_0^{(t)}$} &
      $Q_0^{(s)}$ & \multicolumn{2}{c}{$Q_0^{(t)}$} \\
      \cline{3-4}\cline{6-7} 
      & & $\Delta J=1$ & $\Delta J=2$ & & $\Delta J=1$ & $\Delta J=2$
      \\ \hline 
      $3/2^-$ &  100 & & & 103 &  &  \\ 
      $5/2^-$ &  138 & 87 (266) & & 143 & 95 & \\ 
      $7/2^-$ &  67 & 99 & 88 (80) & 85 & 102 & 95 \\ 
      $9/2^-$ &  101 & 75 & 82 (92) & 105 & 87 & 95 \\      
      $11/2^-$ &  69 & 100 & 87 (89) & 84 & 102 & 98 \\
      $13/2^-$ &  101 & 66 & 77 & 106 & 80 & 87 \\ 
      $15/2^-$ &  69 & 103 & 81 & 83 & 105 & 91 \\ 
      $17/2^-_1$ & $-5.5$ & 22 & 17 & 2.7 & 35 & 18 \\ 
      $17/2^-_2$ & 68 & 41 & 66 & 63 & 57 & 71 \\ 
      $19/2^-$ &  39 & 106, 109 & 65 & 41 & 88, 122 & 73 \\  
      $21/2^-$ &  38 & 35 & 54, 39.6 & 37 & 5.4 & 53, 42 \\ 
      $23/2^-$ &  42 & 43 & 66 & 41 & 26 & 69 \\
      $25/2^-$ &  45 & 53 & 57 & 40 & 8.4 & 59 \\ 
      $27/2^-$ &  38 & 63 & 55 & 28 & 66 & 55 \\ 
      $29/2^-$ &  35 & 22 & 32 & 29 & 13 & 40 \\ 
      $31/2^-$ &  42 & 69 & 41 & 30 & 29 & 48 \\ 
    \end{tabular}
  \end{center}
\end{table}

\begin{table}
  \begin{center}
    \caption{Intrinsic quadrupole moments ($e$ fm$^2$) of the mirror
      pair $^{49}$Cr-$^{49}$Mn. The numbers in parenthesis have been
      calculated using the experimental data. $Q_0^{(s)}$ means
      computed from the spectroscopic moment. $Q_0^{(t)}$ from the
      $B(E2)$ value. When a state can decay to two, both numbers are
      given.}
    \label{tab:q0_49}
    \leavevmode
    \begin{tabular}{ccccccc}
      & \multicolumn{3}{c}{$^{49}$Cr} & \multicolumn{3}{c}{$^{49}$Mn}
      \\ 
      $J^\pi_n$ & $Q_0^{(s)}$ & \multicolumn{2}{c}{$Q_0^{(t)}$} &
      $Q_0^{(s)}$ & \multicolumn{2}{c}{$Q_0^{(t)}$} \\
      \cline{3-4}\cline{6-7} 
      & & $\Delta J=1$ & $\Delta J=2$ & & $\Delta J=1$ & $\Delta J=2$
      \\ \hline 
      $5/2^-$ & 101 & & & 102 &  &  \\
      $7/2^-$ & 142 & 98 (104) & & 114 & 101 &  \\
      $9/2^-$ & 92 & 98 (119) & 100 (122) & 102 & 95 & 101 \\
      $11/2^-$ & 69 & 97 & 101 (112) & 73 & 97 & 102 \\
      $13/2^-$ & 98 & 94 & 96 (55) & 98 & 87 & 95 \\
      $15/2^-$ & 47 & 82 & 88 (61) & 39 & 86 & 85 \\
      $17/2^-$ & 34 & 72 & 75 & 40 & 74 & 74 \\
      $19/2^-_1$ &  9.8 & 87 & 76 (67) & 4.5 & 105 & 74 \\
      $19/2^-_2$ &  50 & 17 & 37 & 37 & 4.1 & 40 \\
      $21/2^-$   &  29 & 67, 30 & 66 & 39 & 76, 32 & 68 \\
      $23/2^-$   &  13 & 87 & 68, 14 & 9.0 & 112 & 72, 6.7 \\
      $25/2^-$   & $-39$ & 8.6 & 17 & $-44$ & 8.4 & 22 \\
      $27/2^-$   & $-5.1$ & 50 & 52 & $-16$ & 34 & 55 \\
      $29/2^-$   & $-1.4$ & 35 & 43 & $-18$ & 12 & 33 \\
      $31/2^-$   & $-4.6$ & 28 & 42 & $-24$ & 33 & 37 \\
    \end{tabular}
  \end{center}
\end{table}

\end{document}